\documentclass[9pt,academicons]{article}

\usepackage{CADA}

\newcommand{\rev}[1]{#1}

\DeclareMathOperator*{\argmin}{arg\,min}

\title{Approximating Triangular Meshes by Implicit, Multi-Sided Surfaces}

\begin{document}

\maketitle

\authorSection{
	\anAuthor{\'Agoston Sipos}{0000-0002-5562-2849}{1},
	\anAuthor{Tam\'as V\'arady}{0000-0001-9547-6498}{2},
	\anAuthor{P\'eter Salvi}{0000-0003-2456-2051}{3} 
}

\affiliationSection{
	\anAffiliation{1}{Budapest University of Technology and Economics}{asipos@edu.bme.hu}
	\anAffiliation{2}{Budapest University of Technology and Economics}{varady@iit.bme.hu}
	\anAffiliation{3}{Budapest University of Technology and Economics}{salvi@iit.bme.hu}
}

\correspondingAuthor{\'Agoston Sipos}{asipos@edu.bme.hu}

\abstract{
The \emph{I-patch} is a multi-sided surface representation, defined as
a combination of implicit \emph{ribbon} and \emph{bounding} surfaces,
whose pairwise intersections determine the natural boundaries of the
patch.  Our goal is to show how a collection of smoothly connected
I-patches can be used to approximate triangular meshes. We start from
a coarse, user-defined vertex graph which specifies an initial
subdivision of the surface. Based on this, we create ribbons that
tightly fit the mesh along its edges in both positional and tangential
sense, then we optimize the free parameters of the patch to better
approximate the interior. If the surfaces are not sufficiently
accurate, the network needs to be refined; here we exploit that the
I-patch construction naturally supports T-nodes. We also describe a
normalization method that nicely approximates the Euclidean distance
field, and can be efficiently evaluated. The capabilities and
limitations of the approach are analyzed through several examples.
}


\doi{10.14733/cadaps.2022.aaa-bbb}

\section{INTRODUCTION}

Implicit surfaces have many nice qualities. They are generally
$C^\infty$-continuous and represent half-spaces, so point-membership
classification is straightforward. No parameterization is needed for
distance computations and approximating data points, and they are
favorable for several geometric interrogations, such as ray tracing or
surface intersections.
Simple regular shapes (such as planes, cylinders, cones, etc.) are
commonly represented by implicit surfaces. Free-form modeling,
however, is more controversial. Counter-arguments include various
shape problems, such as singularities, self-intersections, handling
several disconnected surface portions, and the rigidity of the shapes
using only limited shape parameters.

In this paper we investigate how to approximate a triangular mesh by a
collection of smoothly connected, multi-sided patches. We wish to
preserve the benefits of the implicit representation and try to
eliminate difficulties by retaining the patch within a well-defined
subspace. We assume that the triangular mesh represents an object
bounded by large free-form faces, without high curvature variations or
tiny features. We also assume that there exists a coarse initial
network associated with the mesh; its vertices determine the corners
of the patches to be constructed, and its edges define loops for the
patchwork. Such a network may be created by a user, produced by an
automatic quad meshing algorithm, or defined by a 3D cell
structure---examples will be given later.  We compute an approximating
patchwork and evaluate the deviations. If the accuracy is not
sufficient, we adaptively refine the initial structure by subdivision
(permitting T-nodes) and then refit as needed.

Our representation of choice is the I-patch~\cite{ipatch1}, a class of
implicit surfaces that interpolate a loop of boundary curves. Each
segment is defined as the intersection of a \emph{ribbon} (primary
surface) and a \emph{bounding surface}, both given in implicit form,
similarly to functional splines~\cite{functional}. \rev{Thus the
boundaries may be general 3D curves, or planar curves when either
the ribbon or the bounding surface is represented by a plane.}
I-patches smoothly join to the given ribbons. In this paper we
prescribe only $G^1$ connections, although the algebra permits
higher-degree continuity, as well. The ribbons and bounding surfaces
are determined in such a way that the I-patch interpolates corner
points on the mesh and approximates underlying data points. I-patches
have a few free scalar parameters, by means of which the interior can
be efficiently optimized.

The paper is structured in the following way. After briefly reviewing
previous work in Section~\ref{sec:previous}, we describe the basic
equations of I-patches in Section~\ref{sec:preliminaries}. We
introduce a faithful distance field for I-patches
(Section~\ref{sec:faithful}) and then explain what sort of ribbon and
bounding surfaces (Section~\ref{sec:construction}) will be used for an
optimized mesh approximation (Section~\ref{sec:approximation}). The
concept of refining the curve network will be presented in
Section~\ref{sec:refinement}, followed by a few examples to illustrate
the capabilities and limitations of our approach
(Section~\ref{sec:examples}).

\section{PREVIOUS WORK}
\label{sec:previous}

There is a vast range of algorithms for creating implicit surfaces
interpolating or approximating point data. These can be categorized
into several classes, such as methods based on local
approximations~\cite{Hoppe:1992, Ohtake:2005}, algebraic
splines~\cite{Juttler:2002, Hamza:2020}, radial basis
functions~\cite{Carr:2001, Ohtake:2003}, moving least
squares~\cite{Shen:2004}, Poisson surface
reconstruction~\cite{Kazhdan:2013}, or even neural
networks~\cite{Takikawa:2021}.
All of these methods are very general, and can be used on unorganized
point sets. Because of this, they also share some common properties:
the created function is a highly ``algebraic'' formulation, almost
devoid of geometric meaning, and the resulting surface often has
high-frequency variations.

Our context is much more restricted: the input points are organized
into a mesh, and additional structure is supplied by a user-defined
vertex graph. These allow us to create a model by connecting
relatively low-degree, high-quality \emph{single-patch} surfaces
composed of geometrically intelligible parts.
In this sense, implicit modeling techniques are more relevant to this
research, including the blending methods of
Rockwood~\cite{Rockwood:1989} and Warren~\cite{Warren:1992}, as well
as A-patches~\cite{Bajaj:1995} and functional
splines~\cite{functional,Hartmann:2001}, although these were not used
for approximation \emph{per se}.

In this work we investigate the approximation capabilities of
I-patches~\cite{ipatch1}. This is a continuation of our former
research, where we have shown how to construct I-patches for
applications in polyhedral design and setback vertex blending.
The I-patch representation is inherently multi-sided, with implicitly
defined, but exact boundary curves and cross-derivative
constraints. This sets it apart from the other blending methods
outlined above---see our previous paper~\cite{ipatch2} for more
information and some comparisons.

\section{PRELIMINARIES: I-PATCHES}
\label{sec:preliminaries}

\begin{figure}[!ht]
  {
    \begin{subfigure}{0.32\textwidth}
      \centering
      \includegraphics[width = \textwidth]{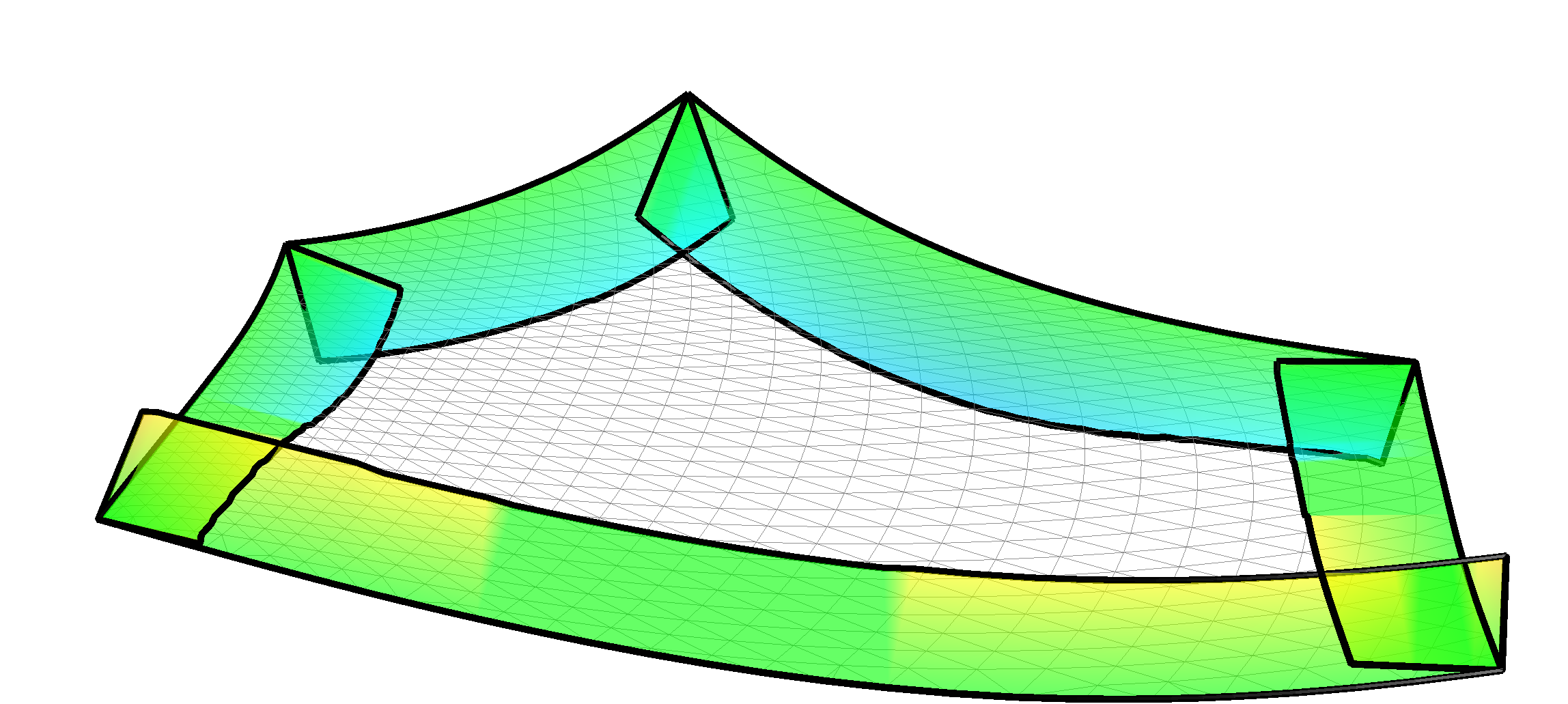}
      \caption{Ribbons}
      \label{fig:construction-ribbons}
    \end{subfigure}
    \hfill
    \begin{subfigure}{0.32\textwidth}
      \centering
      \includegraphics[width = \textwidth]{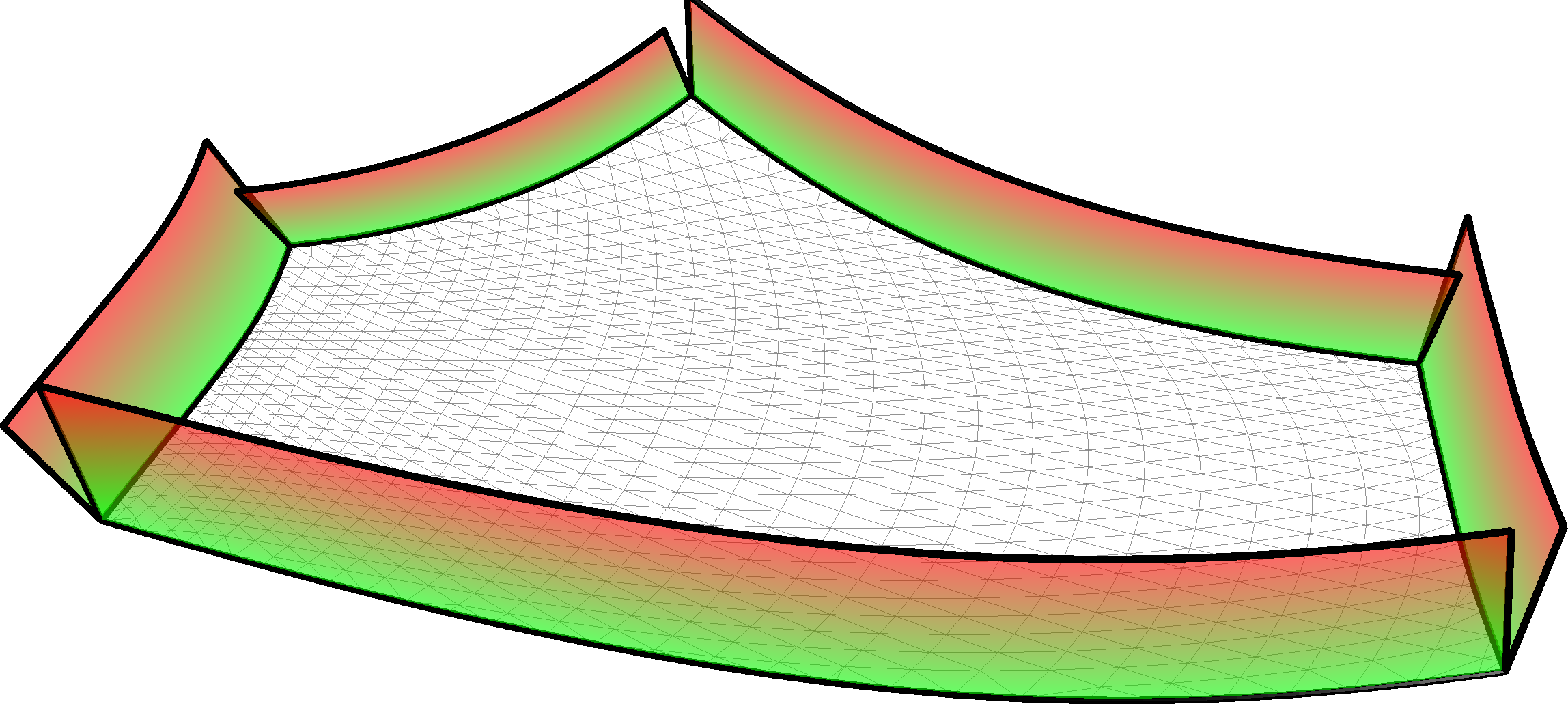}
      \caption{Bounding surfaces}
      \label{fig:construction-boundings}
    \end{subfigure}
    \hfill
    \begin{subfigure}{0.32\textwidth}
      \centering
      \includegraphics[width = \textwidth]{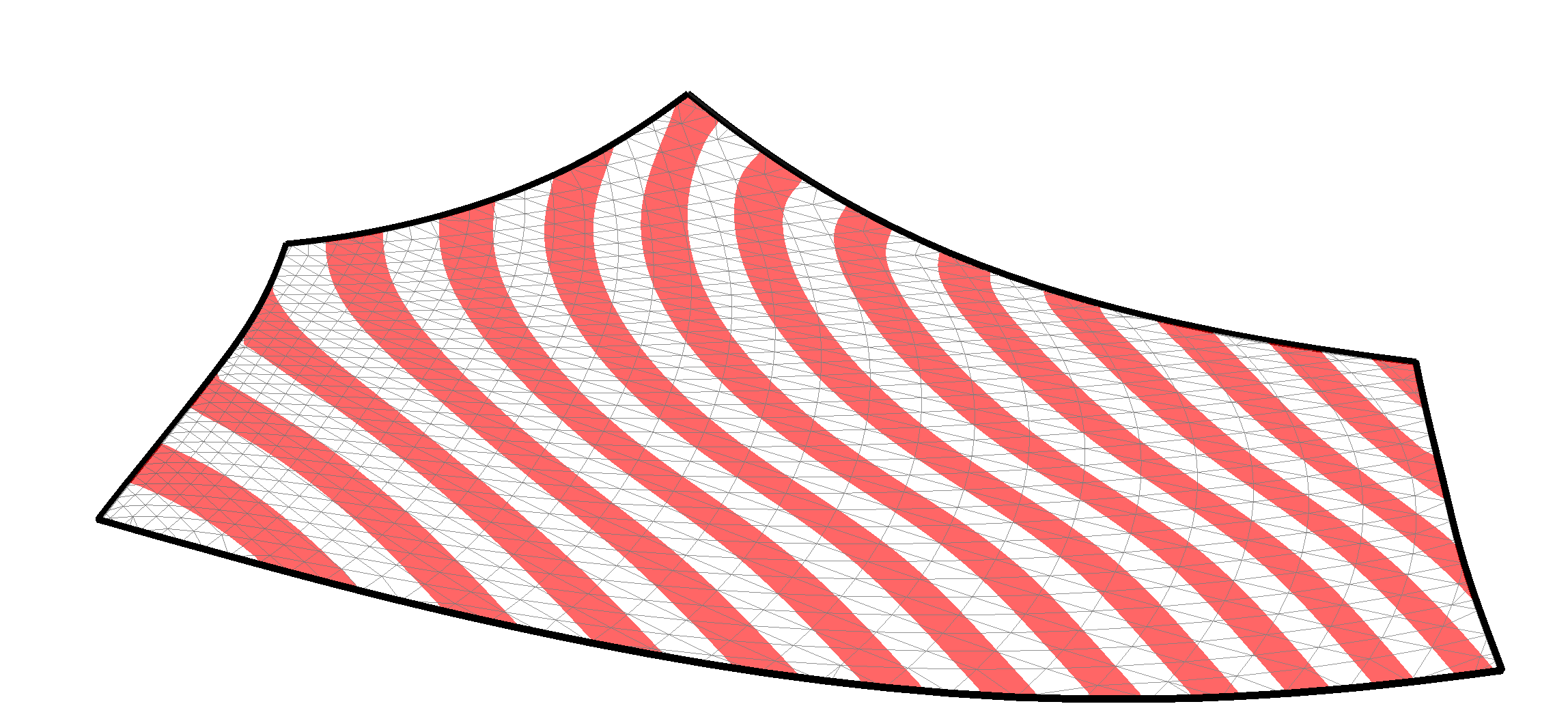}
      \caption{I-patch}
      \label{fig:construction-Ipatch}
    \end{subfigure}
  }
  \caption{Construction of an I-patch.}
  \label{fig:construction}
\end{figure}

An $n$-sided I-patch is defined by $n$ ribbons and bounding surfaces,
given in implicit form as $R_i(x,y,z)=0$ and $B_i(x,y,z)=0$,
respectively ($i=1\dots n$), see Figure~\ref{fig:construction}. The
patch itself is constructed as the 0-isosurface of
\begin{equation}
  \label{eq:ipatch}
  I(x,y,z)=\sum_{i=1}^n w_iR_i(x,y,z)\prod_{j\neq i}B_j^2(x,y,z)
  -w_0\prod_{i=1}^nB_i^2(x,y,z),
\end{equation}
where $w_i$ ($i=0\dots n$) are free scalar parameters. It is defined
within the volume enclosed by the bounding surfaces, i.e., for points
where $B_i(x,y,z)\ge0$ for all $i$. (For readability, we will omit the
$(x,y,z)$ arguments of all implicit surfaces from now on.)

The intersection of $R_i$ and $B_i$ defines the $i$-th boundary
curve, which is interpolated by the I-patch, since all terms of the
above expression vanish there. Similarly, it is
easy to show that at these points the gradient of $R_i$ will be
parallel to the gradient of the I-patch, so $G^1$ continuity is
ensured. (Raising the degrees in
Eq.~(\ref{eq:ipatch}) would guarantee a higher order of continuity.)
When two or more bounding surfaces coincide, special handling is
needed~\cite{ipatch2}.

Apart from points on the bounding surfaces, an equivalent formulation
is given by the $w_0$-isosurface of
\begin{equation}
  \label{eq:ipatch-distance}
  I_{w_0}=\sum_{i=1}^n \frac{w_iR_i}{B_i^2}=:\sum_{i=1}^nd_i,
\end{equation}
where each $d_i$ can be regarded as a distance measure associated with
the $i$-th ribbon, so I-patches can be interpreted as the locus of
3D points where the sum of $n$ distances is constant. This is similar
to the logic of how some classic surfaces, such as ellipsoids, are
derived. We remark that an ellipsoid can in fact be exactly
represented in the above rational form.

\section{FAITHFUL DISTANCE FIELD}
\label{sec:faithful}

The computation of distances from a given surface, or the creation of
an offset surface by a given distance, are heavily used operations in
several applications, such as design, geometric intersections and
interrogations, and data approximation. One advantage of implicit
surfaces is that they possess a natural distance field from their
$0$-isosurfaces. Unfortunately algebraic distances, even when
multiplied by a carefully chosen scalar factor, do not yield Euclidean
distances, except for planes.

A frequently applied, practical method for enhancing distance fields
is to normalize the implicit equation by the norm of its
gradient~\cite{Taubin:1994}. This is a good solution to obtain
approximate Euclidean distances in the vicinity of the $0$-isosurface,
but these expressions contain square roots, and may exhibit
singularities. They also deviate from the correct distance when we move
farther away from the implicit surface.

We propose a normalization for I-patches that produces a good
approximation of the Euclidean distance field, not only in the
vicinity of the surface, but in a much wider range.  The literature
often uses the terminology ``faithful'' for such
distances~\cite{Lukacs:1998}, and we also retain this adjective.

\begin{figure}[!ht]
  {
    \hfill
    \begin{subfigure}{0.49\textwidth}
      \centering
      \includegraphics[width = \textwidth]{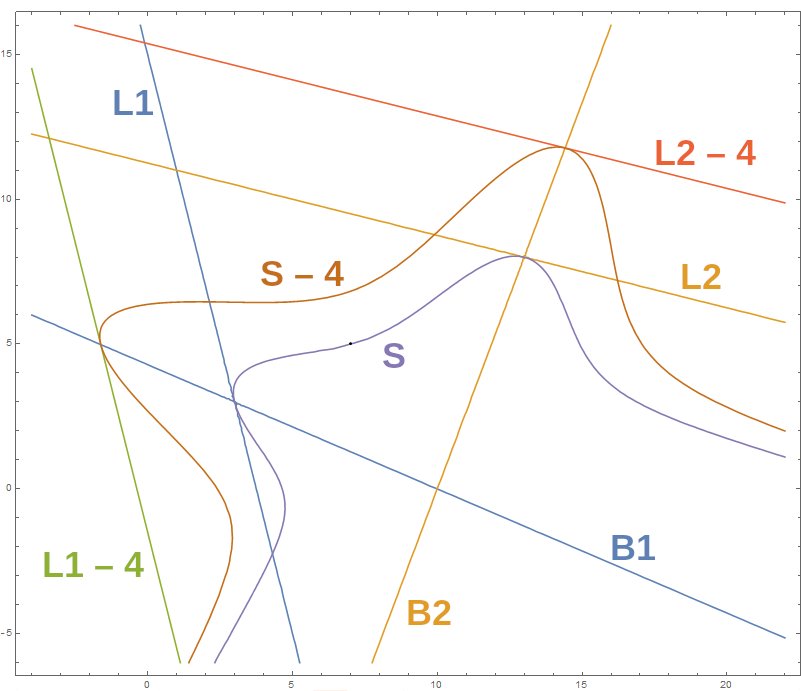}
      \caption{Faithful normalization}
      \label{fig:faithful-a1}
    \end{subfigure}
    \hfill
    \begin{subfigure}{0.49\textwidth}
      \centering
      \includegraphics[width = \textwidth]{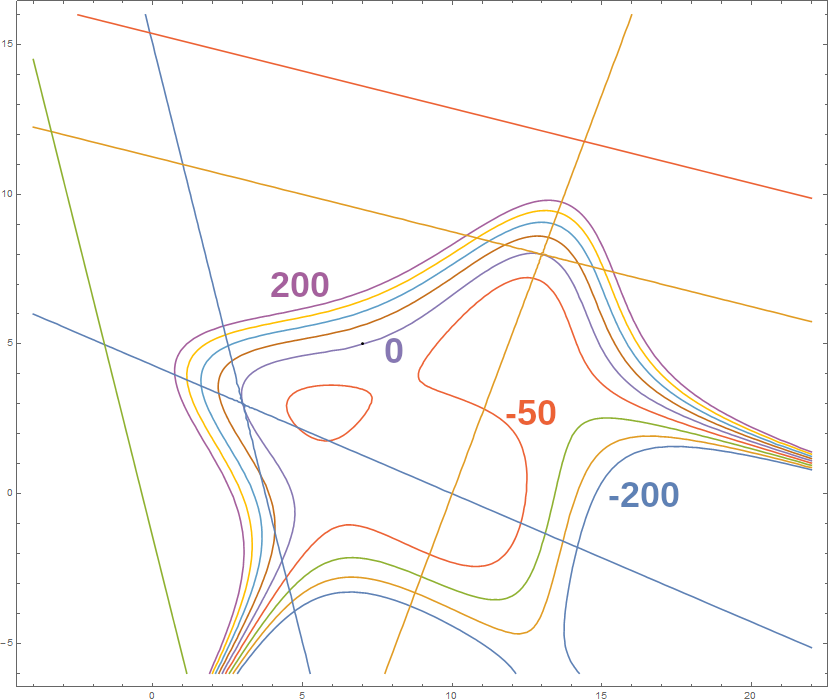}
      \caption{No normalization}
      \label{fig:faithful-a2}
    \end{subfigure}
    \hfill
  }

  {
    \hfill
    \begin{subfigure}{0.49\textwidth}
      \centering
      \includegraphics[width = \textwidth]{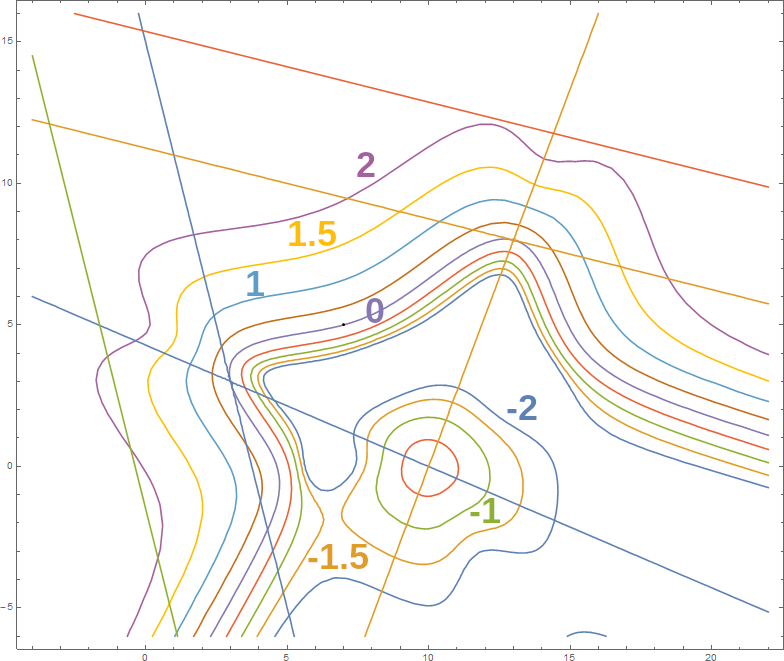}
      \caption{Gradient normalization}
      \label{fig:faithful-a3}
    \end{subfigure}
    \hfill
    \begin{subfigure}{0.49\textwidth}
      \centering
      \includegraphics[width = \textwidth]{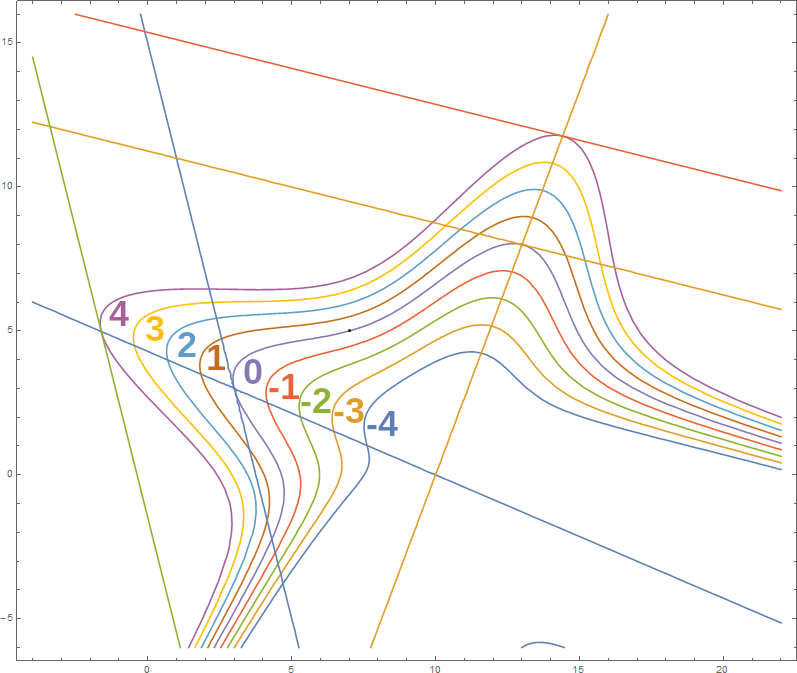}
      \caption{Faithful normalization}
      \label{fig:faithful-a4}
    \end{subfigure}
    \hfill
  }
  \caption{Offsets of an I-segment using different normalizations.}
  \label{fig:faithful-a}
\end{figure}

Using the notation $\alpha_i=w_i/B_i^2$, we can formulate the equation
of the I-patch as a weighted sum of ribbon surfaces $R_i$, multiplied
by blending functions:
\begin{equation}
  \label{eq:faithful1}
  I=\sum_{i=1}^n R_i\cdot\alpha_i -w_0.
\end{equation}
We derive a faithful normalization by dividing with the sum of the
blending functions:
\begin{equation}
  \label{eq:faithful2}
  \hat{I}=\frac{\sum_{i=1}^n R_i\cdot\alpha_i -w_0}
      {\sum_{i=1}^n \alpha_i}.
\end{equation}
Examining an algebraic offset of the normalized I-patch, we find that
\begin{equation}
  \label{eq:faithful3}
  \hat{I} - d=\frac{\sum_{i=1}^n (R_i - d)\cdot\alpha_i-w_0}
      {\sum_{i=1}^n \alpha_i}.
\end{equation}
In other words, the offset of the normalized I-patch is also a
normalized I-patch, created from ribbon offsets, and blended by the
same $\alpha_i$ functions (i.e., the same bounding surfaces).

The faithfulness of this formulation depends on the distance field of
the ribbons. I-lofts (see Section~\ref{sec:construction}), for
example, can be made faithful in this sense, since they are defined by
combining planar ribbons. As a consequence, an I-patch combining
I-lofts will also behave in a faithful manner.

\rev{First we demonstrate this concept in 2D using an I-segment
$S$, which is a planar implicit curve, that blends} together two
lines $L_1$ and $L_2$ (Fig.~\mbox{\ref{fig:faithful-a1}}). We wish to compute
an offset of the I-segment that smoothly connects the accurately
displaced offset lines $L_1-d$ and $L_2-d$, and expect to obtain a
good distance field between them. Figure~\ref{fig:faithful-a2} shows
the distribution of algebraic distances, and
Figure~\ref{fig:faithful-a3} shows the distance field after gradient
normalization, where one can observe the uneven and unproportional
distribution of the offset curves.  Our proposed normalization in
Figure~\ref{fig:faithful-a4} shows a faithful distance field.

Faithful distances are important in our context. Offsets of I-patches
or I-segments can be directly computed yielding good approximations of
Euclidean distances. Approximating data points using I-patches or
I-lofts becomes easy and computationally efficient, since we only need
to substitute into the above normalized equations, and optimize
accordingly.

\begin{figure}[!ht]
  {
    \hfill
    \begin{subfigure}{0.4\textwidth}
      \centering
      \includegraphics[width = \textwidth]{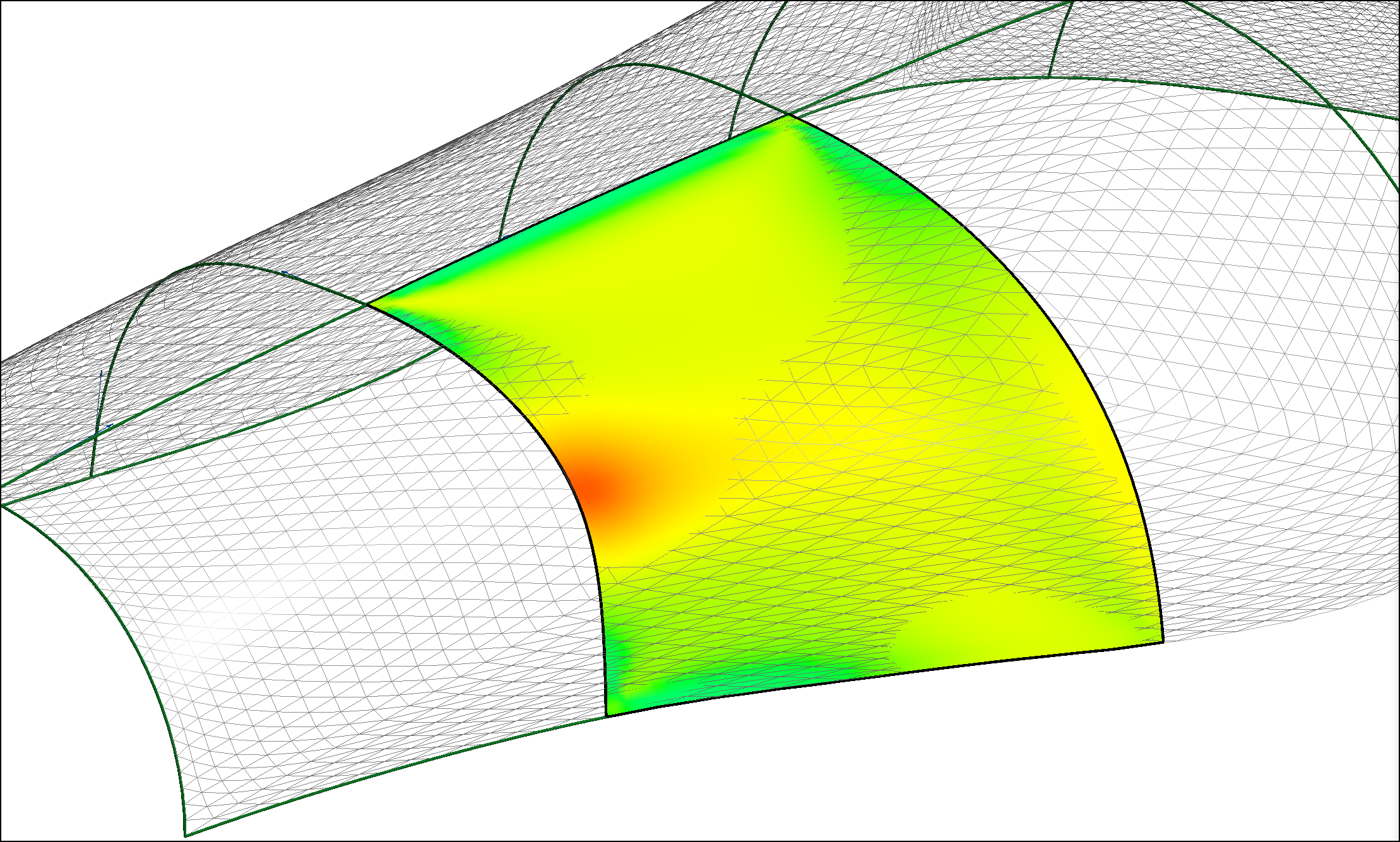}
      \caption{Original surface}
      \label{fig:faithful-b1}
    \end{subfigure}
    \hfill
    \begin{subfigure}{0.4\textwidth}
      \centering
      \includegraphics[width = \textwidth]{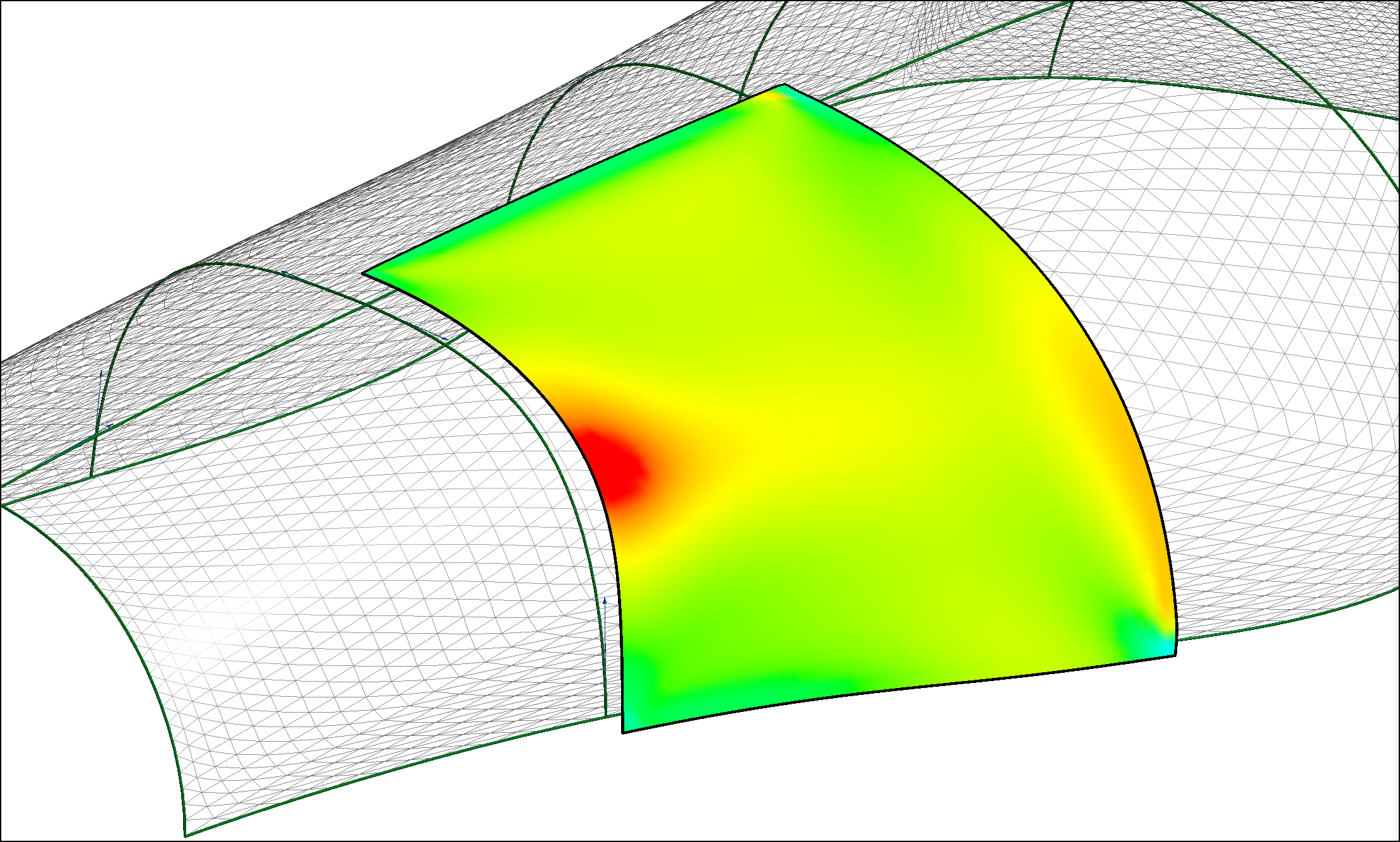}
      \caption{Offset surface}
      \label{fig:faithful-b2}
    \end{subfigure}
    \hfill
  }
  \caption{Offsetting an I-patch.}
  \label{fig:faithful-b}
\end{figure}

\begin{figure}[!ht]
  {
    \hfill
    \includegraphics[width = .55\textwidth]{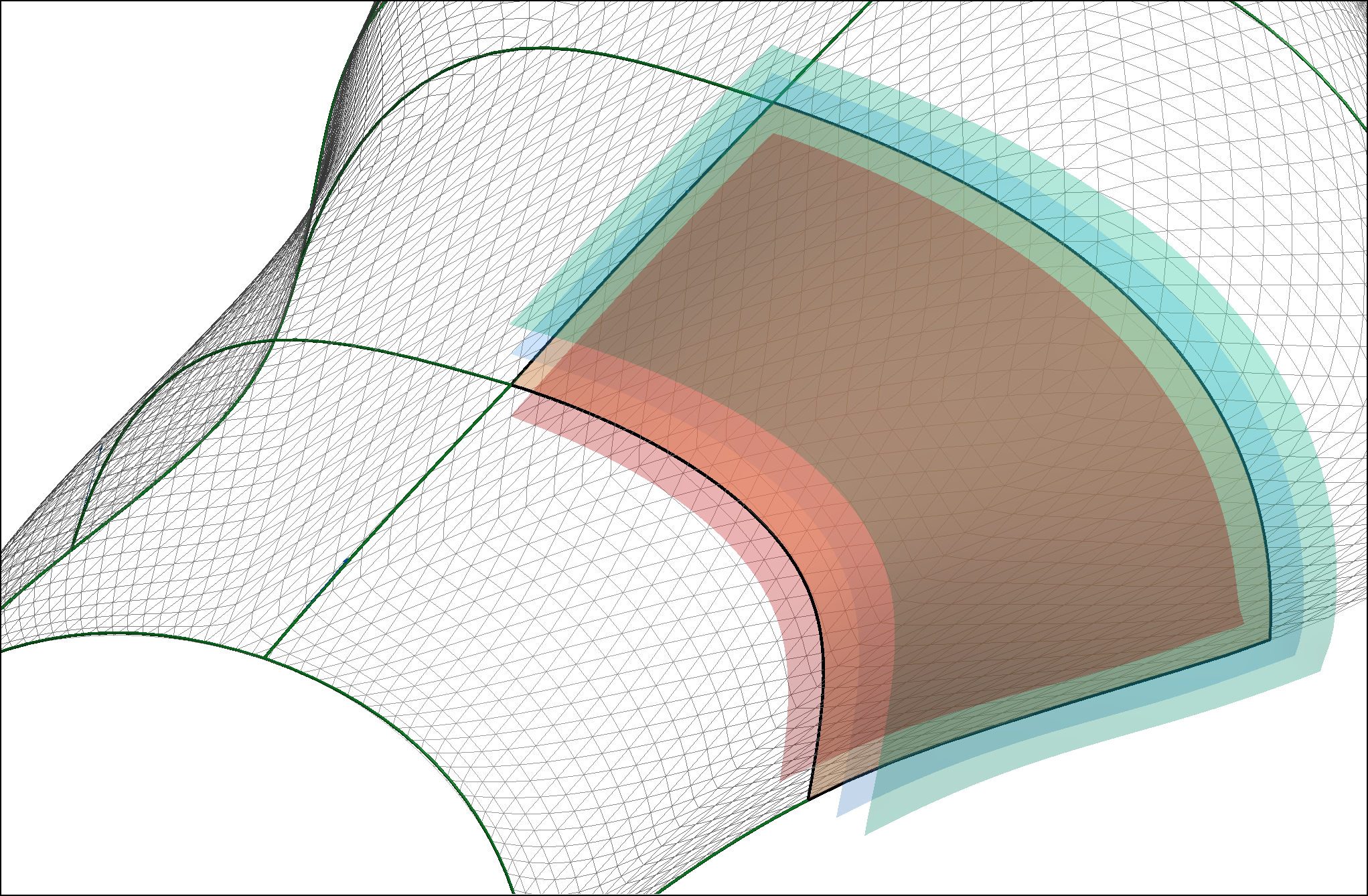}
    \hfill
  }
  \caption{Multiple offsets of an I-patch.}
  \label{fig:faithful-c}
\end{figure}

In Figure~\ref{fig:faithful-b} an I-patch is shown with its offset,
using mean curvature maps. Figures~\ref{fig:faithful-b1}
and~\ref{fig:faithful-b2} show the patch and its offset isosurface,
and Figure~\ref{fig:faithful-c} shows four superimposed offset patches
with $d=-5$, $0$, $5$ and $10$.

Finally, note that the same principle works with the polynomial form
of the I-patch:
\begin{equation}
  \label{eq:faithful4}
  \hat{I}=\frac{\sum_{i=1}^n w_iR_i\prod_{j\neq i}B_j^2
                -w_0\prod_{i=1}^nB_i^2}
      {\sum_{i=1}^n w_i\prod_{j\neq i}B_j^2}.
\end{equation}
This is a more complex, but equivalent equation, which can be
evaluated on the boundary, as well.

\section{RIBBONS AND BOUNDING SURFACES}
\label{sec:construction}

In this section we introduce the ribbons and bounding surfaces needed
for the current I-patch construction. As will be explained below,
these surfaces are always defined as \emph{weighted combinations} of
planes given at the corner points and auxiliary planes derived from
the corner data.

Let us take two adjacent corner points $\mathbf{p}_1$, $\mathbf{p}_2$,
and incident \emph{corner planes} $\pi_1$, $\pi_2$ with normal vectors
$\mathbf{n}_1$, $\mathbf{n}_2$. The corresponding ribbon interpolates
both points and normal vectors. Similarly, the bounding surface
interpolates both points, but it is transversal to the ribbon
surface. The ribbons and bounding surfaces are described by three
kinds of equations: (i)~planes, (ii)~\emph{Liming-surfaces}, and
(iii)~\emph{I-lofts}.

\begin{figure}[!ht]
  {
    \hfill
    \begin{subfigure}{0.45\textwidth}
      \centering
      \includegraphics[width = \textwidth]{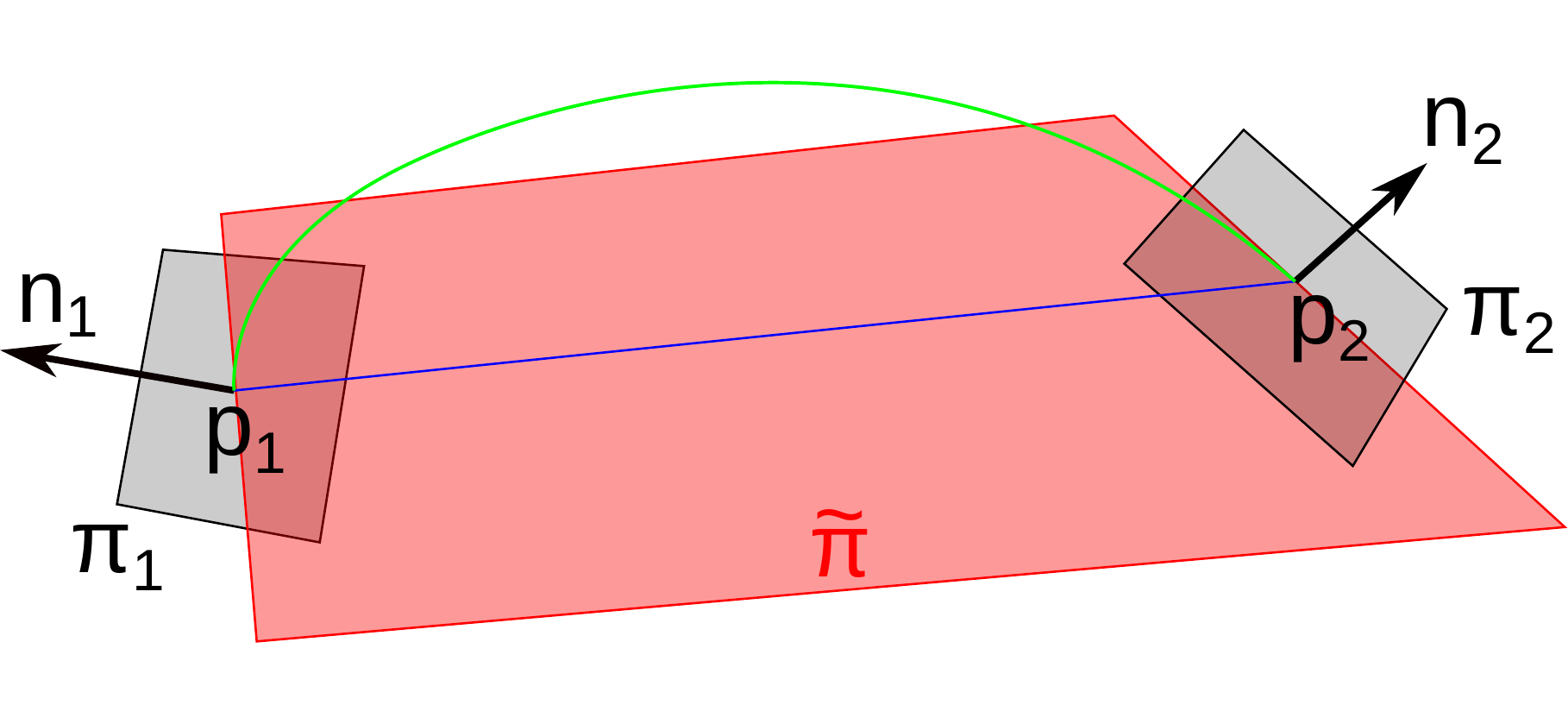}
      \caption{Liming ribbon}
      \label{fig:construction-liming}
    \end{subfigure}
    \hfill
    \begin{subfigure}{0.45\textwidth}
      \centering
      \includegraphics[width = \textwidth]{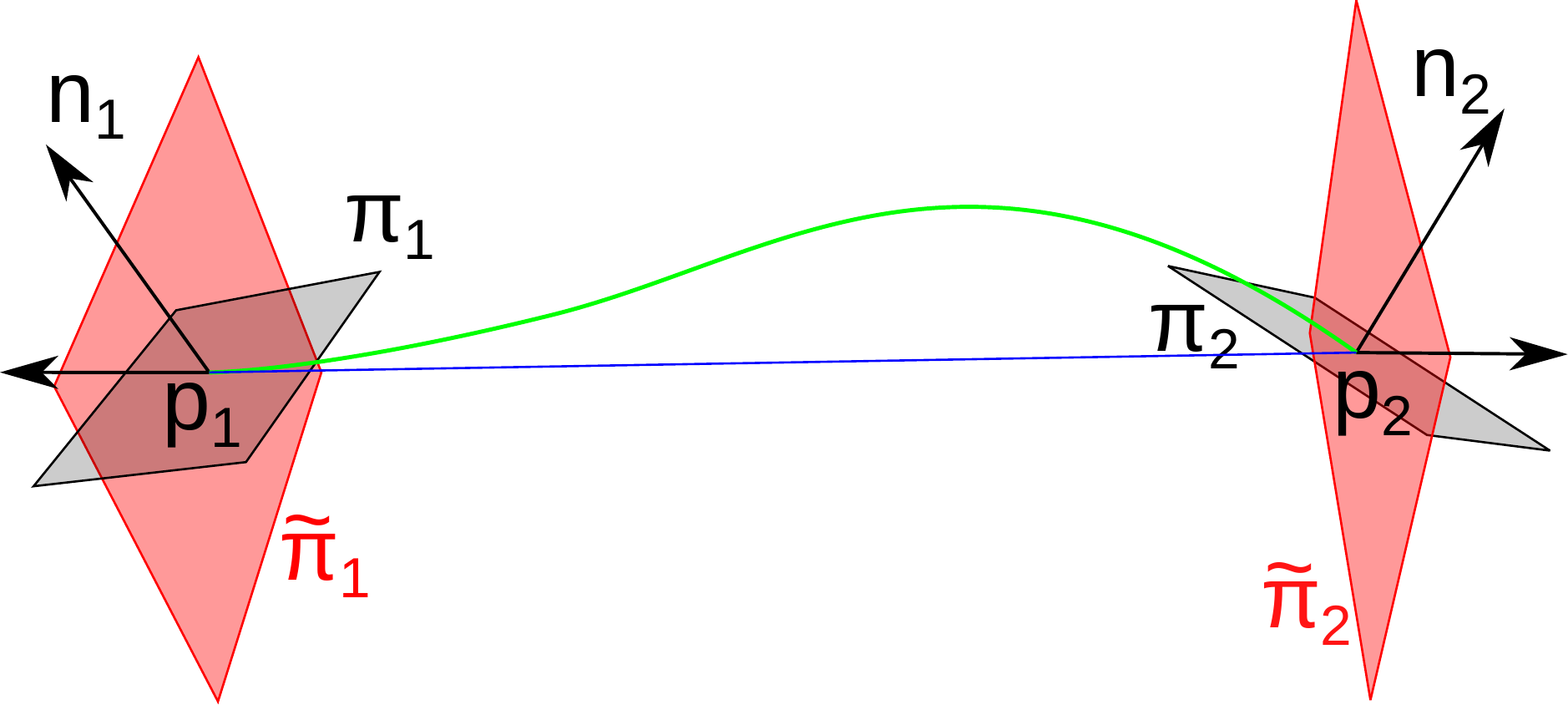}
      \caption{I-loft ribbon}
      \label{fig:construction-iloft}
    \end{subfigure}
    \hfill
  }
  \caption{Constructing ribbons.}
  \label{fig:construction-rib}
\end{figure}

Liming~\cite{liming} created conic curves as a combination of three
implicitly given lines. A Liming-surface, in turn, is the combination
of three planes given in implicit form. In addition to the corner
planes $\pi_i$, an auxiliary cutting plane $\tilde\pi$ is needed that
contains both $\mathbf{p}_1$ and $\mathbf{p}_2$ (see
Fig.~\ref{fig:construction-liming}).  The surface equation is given as
\begin{equation}
  (1-\lambda)\pi_1\pi_2-\lambda\tilde\pi^2=0,
\end{equation}
where $\lambda$ controls the fullness. \rev{The choice of the cutting
plane $\tilde\pi$ does affect the shape of the Liming-surface.
Nevertheless, the simplest setting of its normal vector
$\mathbf{n}_{12}=(\mathbf{n}_1+\mathbf{n}_2)/2$ proved to be
satisfactory. The parameter $\lambda$ is going to be defined by the
optimal approximation of the underlying polyline boundary, see the
next section.}

I-lofts are simple, two-sided I-patches defined by two pairs of planes
containing the corner points. In this case primary ribbons correspond
to the tangent planes at the corners ($\pi_i$), and the local bounding
planes $\tilde\pi_i$ are transversal to these (see
Fig.~\ref{fig:construction-iloft}).  The equation of the I-loft is
\begin{equation}
  w_1\pi_1\tilde\pi_2^2+w_2\pi_2\tilde\pi_1^2-\tilde\pi_1^2\tilde\pi_2^2=0.
\end{equation}
There is freedom to select the bounding planes, as will be explained
below.  The weights $w_1$ and $w_2$ are shape parameters to set the
fullness of I-lofts, \rev{and thus ensure the best approximation of the
underlying polyline boundary, see the next section.}

In our context we prefer to use \emph{planar bounding surfaces}
to subdivide the interior of the mesh, since these lead to planar
boundary curves, and the equation of the I-patch becomes
simpler. Intersections with Liming ribbons yield conic curves, while
we get planar I-segments with I-lofts.  One simple formula to set the
normal of this bounding plane comes from averaging the corner normals:
\begin{equation}
  \mathbf{n}_B=\mathbf{n}_{12}\times(\mathbf{p}_2-\mathbf{p}_1)
  \text{, where }\mathbf{n}_{12}=\frac{\mathbf{n}_1+\mathbf{n}_2}{2}.
\end{equation}
Alternatively, we may rotate the bounding plane around the chord
$\mathbf{p}_2-\mathbf{p}_1$, and may set it to contain a prescribed
mesh point. It is a natural assumption that an optimal ribbon normal
at the midpoint of the boundary curve should approximate the normal
vector of the mesh.

\rev{\emph{Curved} bounding surfaces may be used in the interior of
the mesh to approximate non-planar subdividing curves, but they
\emph{must} be used for the approximation of curved external
boundaries.}  In this case, we calculate the local direction of the
mesh boundary at the corner points; the tangent planes of the bounding
surface are defined by this vector and the corresponding mesh
normal. When a curved bounding surface is intersected with a curved
ribbon, a general 3D curve is obtained.

It should be noted, that although we always attempt to use the
simplest ribbon--bounding surface pair, the use of Liming-surfaces is
not always possible. If the boundary to be approximated has a point of
inflection, or the corner normals are twisted, only I-lofts can be
used~\cite{ipatch2}.

It is also important to observe that the ribbons and the bounding
surfaces depend only on the corner points, the corner normals and the
mesh, so if two adjacent I-patches share the same
$\mathbf{p}_1\mathbf{p}_2$ boundary, the identical surface components
will guarantee a smooth connection.

\section{I-PATCHES TO APPROXIMATE MESHES}
\label{sec:approximation}

In this section we describe how to construct I-patches that
approximate a given mesh.  At this point we assume that the bounding
surfaces have already been determined.  We intersect these with the
mesh, and trace polylines between the adjacent corner points.  We do
not need tracing along the external edges of the mesh, since there the
polylines are directly available. First we create ribbons that
approximate these polylines, then an I-patch that approximates the
interior points of the mesh.

Ribbons match the given corner points and the corresponding
normals \rev{(approximated from the mesh with standard
methods~\mbox{\cite{Max:1999,Cazals:2005}})}. We construct
a Liming-ribbon by optimizing its free $\lambda$
parameter.  This sets the fullness of the ribbon, and determines the
best approximating conic boundary.  I-lofts are somewhat more
flexible, and we have two independent scalar weights for the
optimization. The error function to be minimized is the squared sum of
the faithful algebraic distances\footnote{Alternatively, we also made
experiments with Euclidean distances, but there was no substantial
difference, and the algebraic method was much faster.}  (see
Section~\ref{sec:faithful}).  \rev{As mentioned earlier, it may be
possible that a boundary can be approximated by both a Liming-surface
and an I-loft. In this case, we choose the one with smaller error.}

\begin{figure}[!ht]
  {
    \hfill
    \includegraphics[width = .45\textwidth]{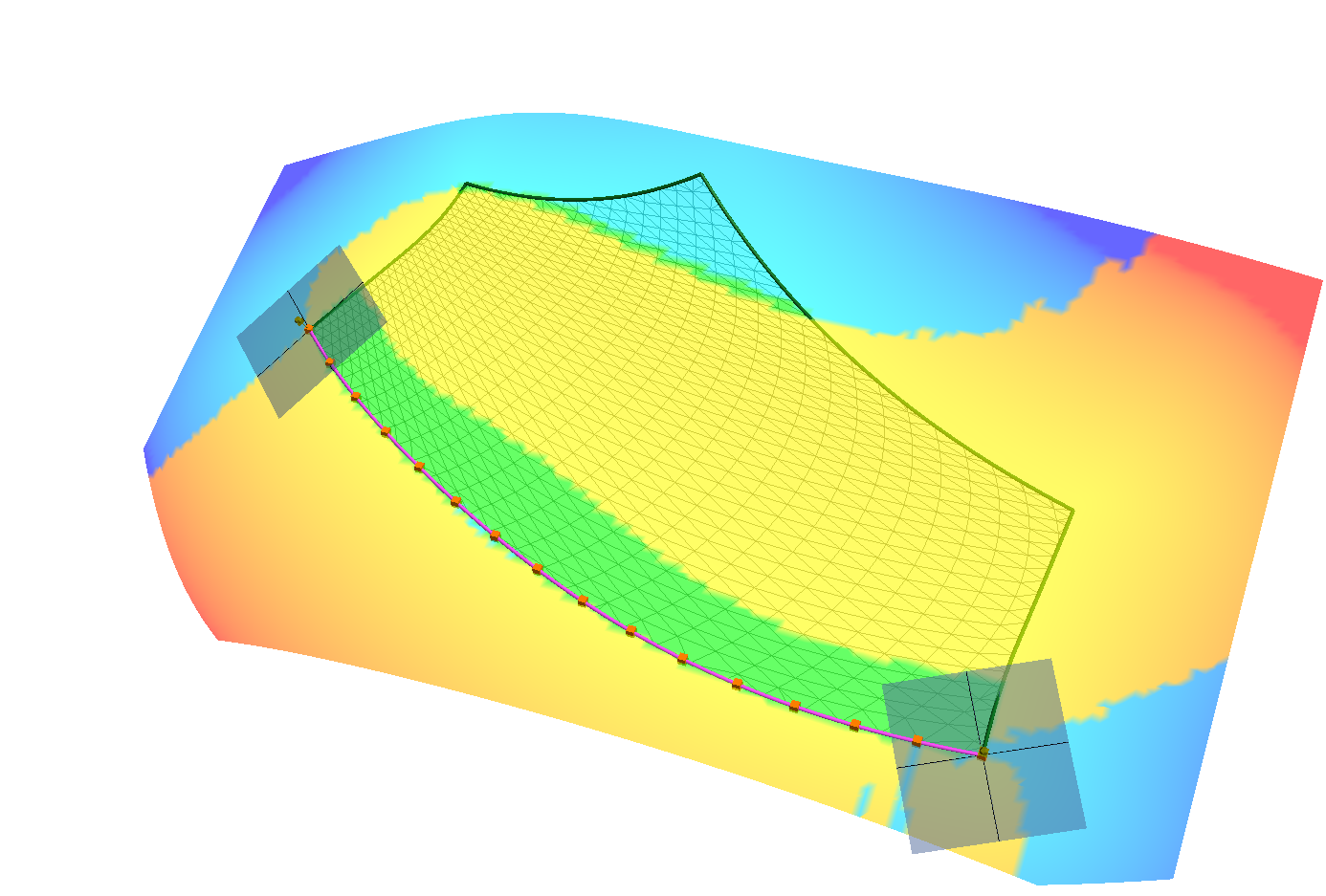}
    \hfill
    \includegraphics[width = .45\textwidth]{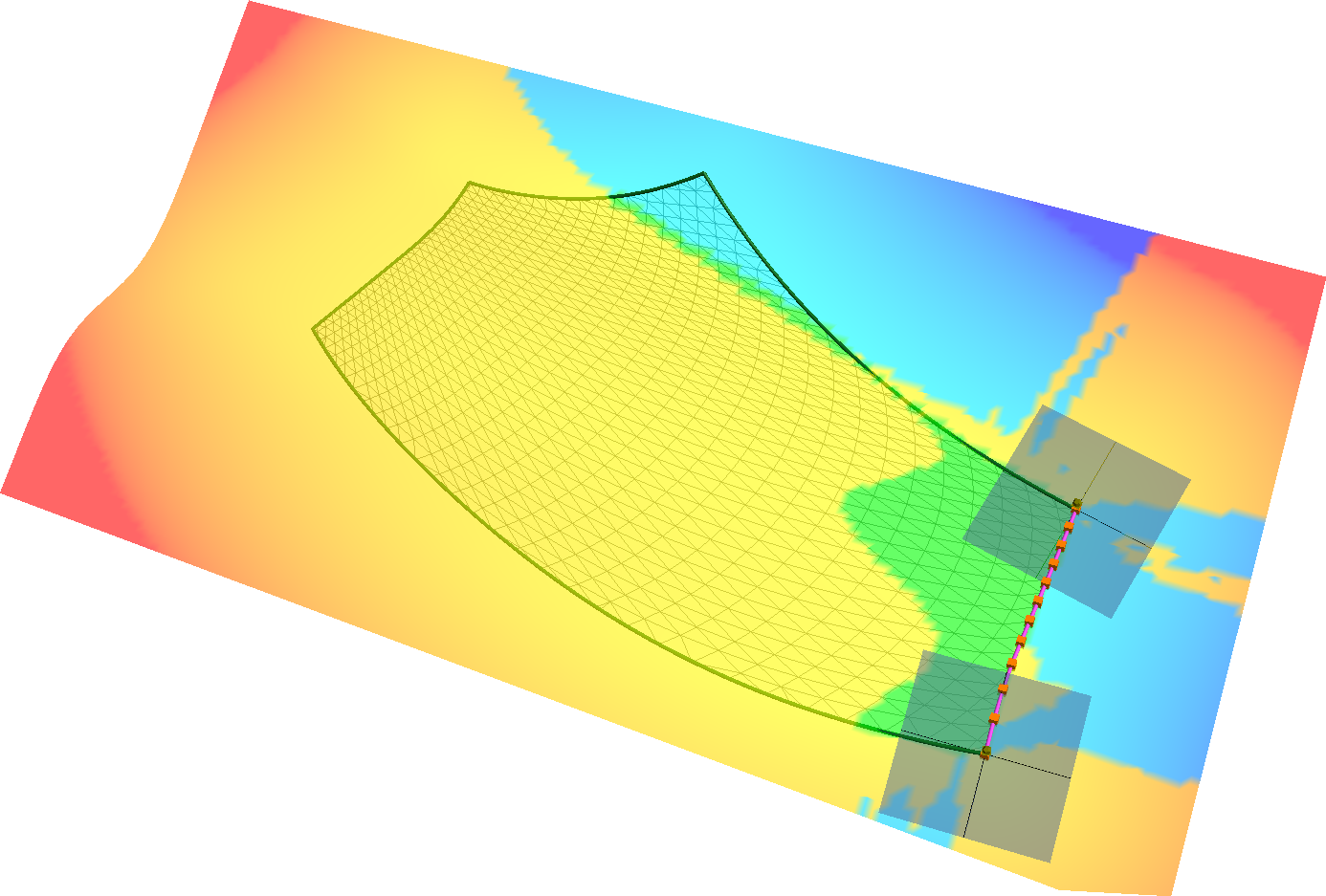}
    \hfill
  }
  \caption{A Liming (left) and an I-loft ribbon (right), showing deviation from the mesh.}
  \label{fig:ribbons}
\end{figure}

Figure~\ref{fig:ribbons} shows a Liming and an I-loft ribbon, with
corner planes and sampled points on the mesh. The deviation map
indicates that the ribbons are close to the mesh (green) in the
vicinity of the related boundary.

Next, mesh points are filtered out so only those remain that
are enclosed by the bounding surfaces. Then we compute the mass center
of the polyline boundaries of the patch, and project this point onto
the mesh. This initial center point $\mathbf{c}$ is used to determine
the default weights $w_i$ for the ribbons by enforcing that all terms
$w_i/B^2_i(\mathbf{c}), i=1,..,n$ are equal. In other words, the
default patch is composed in a way that the individual ribbons
contribute to the patch in a uniform manner at the center point; the
patch is also constrained to interpolate point $\mathbf{c}$. This
initial setting can be iteratively improved to obtain the best
approximation of the underlying data. We minimize the squared sum of
the faithful distances, and set the optimal weights accordingly.
For both the ribbon and patch optimization, derivative-free methods
can be applied; we have chosen the Nelder--Mead
algorithm~\cite{optimization}.

We have observed that it is a good idea to limit the ratios of the
optimized weights. Otherwise one ribbon may dominate the patch
equation, and corrupt the surface quality along the other
borders. Accordingly, we constrained how much the weights of the
optimized ribbons can deviate from their initial values. \rev{This
maximum ratio is a user parameter $\omega$. (In our examples,
$\omega$ was set to $5$.)
}

\begin{figure}[!ht]
  {
    \begin{subfigure}{0.45\textwidth}
      \centering
      \includegraphics[width = \textwidth]{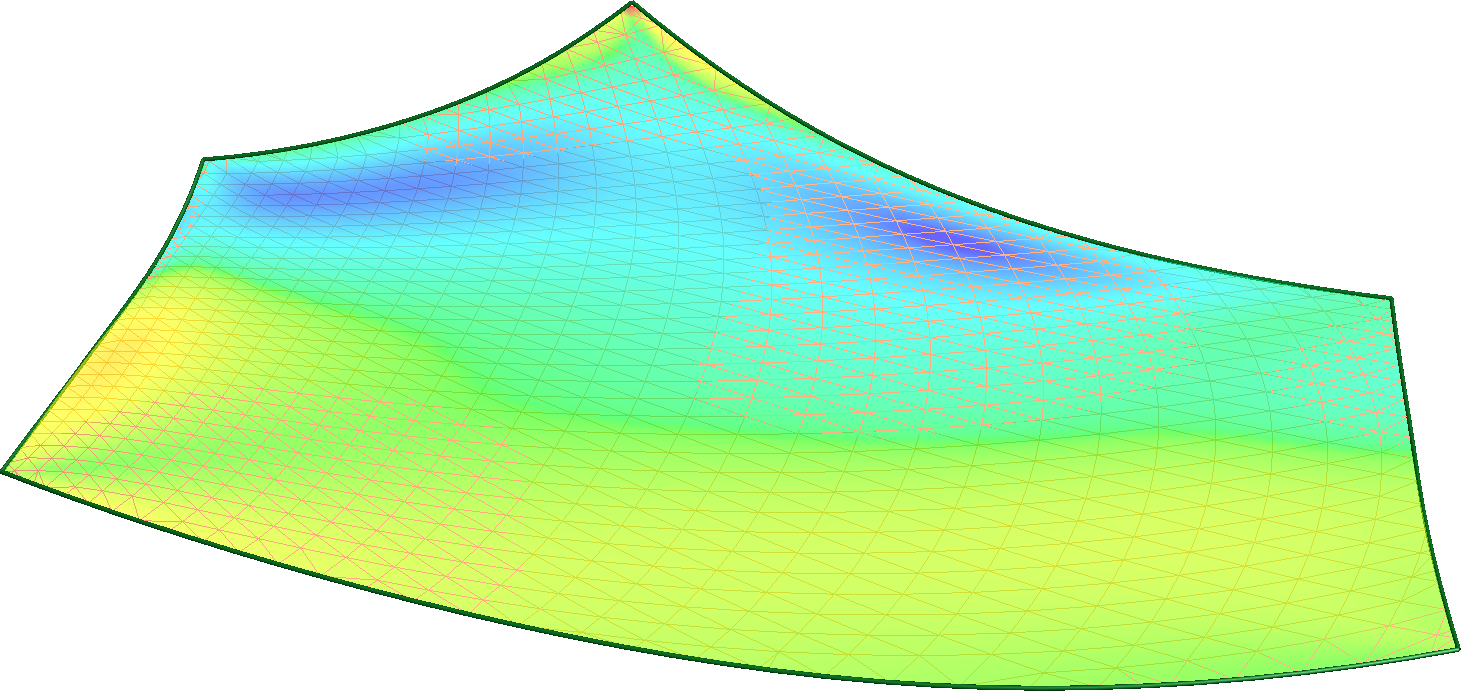}
      \caption{Mean curvature map}
      \label{fig:curvature}
    \end{subfigure}
    \hfill
    \begin{subfigure}{0.45\textwidth}
      \centering
      \includegraphics[width = \textwidth]{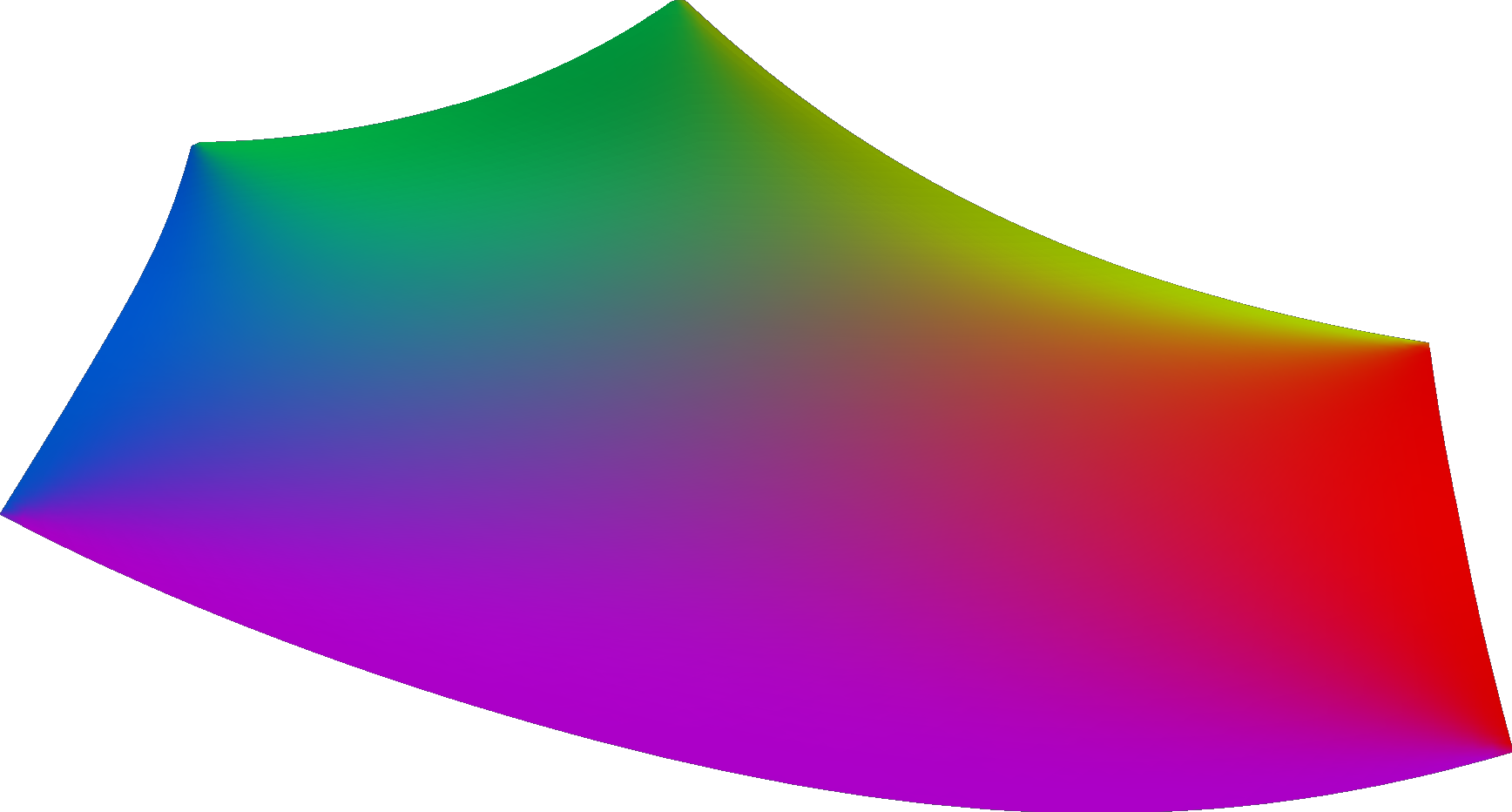}
      \caption{Contributions of each side}
      \label{fig:colors}
    \end{subfigure}
  }
  \caption{Optimization of a 5-sided patch.}
  \label{fig:optimizations}
\end{figure}

\rev{The formalized optimization problem is:}
\begin{equation}
  \argmin_\mathbf{w} \sum_{i=1}^N I_\mathbf{w}(\mathbf{q}_i),\text{ s.t. }
  \max_{i,j} \left(\frac{w_i R_i(\mathbf{c})}{B_i^2(\mathbf{c})} \bigg/
  \frac{w_j R_j(\mathbf{c})}{B_j^2(\mathbf{c})}\right) \leq \omega,
\end{equation}
\rev{where $\mathbf{w} = \{w_i\}_1^n$ is the vector of weights, and
$\mathbf{q}_i$ ($i=1\dots N$) denotes the data points to be approximated.}
Figure~\ref{fig:curvature} shows the curvature map of the optimized
I-patch from Figure~\ref{fig:construction-Ipatch}. We have assigned
colors to the individual sides, and show the strength of the
normalized blending functions in Figure~\ref{fig:colors}.

\section{ADAPTIVE REFINEMENT OF THE PATCHWORK}
\label{sec:refinement}

Once an initial patchwork has been generated, we check whether it
approximates the data points within a prescribed tolerance and---if
necessary---we perform an adaptive refinement following simple
heuristic rules. If a boundary is out of tolerance, it will be
subdivided halfway between its endpoints. If a patch is out of
tolerance, central splitting will be applied and a new mesh point will
be inserted in its interior.  We connect the new subdivision points,
and add new artificial points in order to create a consistent
structure. This leads to a new graph of edges with new ribbons and
bounding surfaces, \rev{given in the same form as before, i.e.,
segments defined by two endpoints with normals, and an underlying
polyline to be approximated. The above procedure automatically
iterates until the requested accuracy is achieved. We remark that
the initial network affects the structure of the final patchwork.}

\begin{figure}[!ht]
  {
    \hfill
    \begin{subfigure}{0.4\textwidth}
      \centering
      \includegraphics[width = \textwidth]{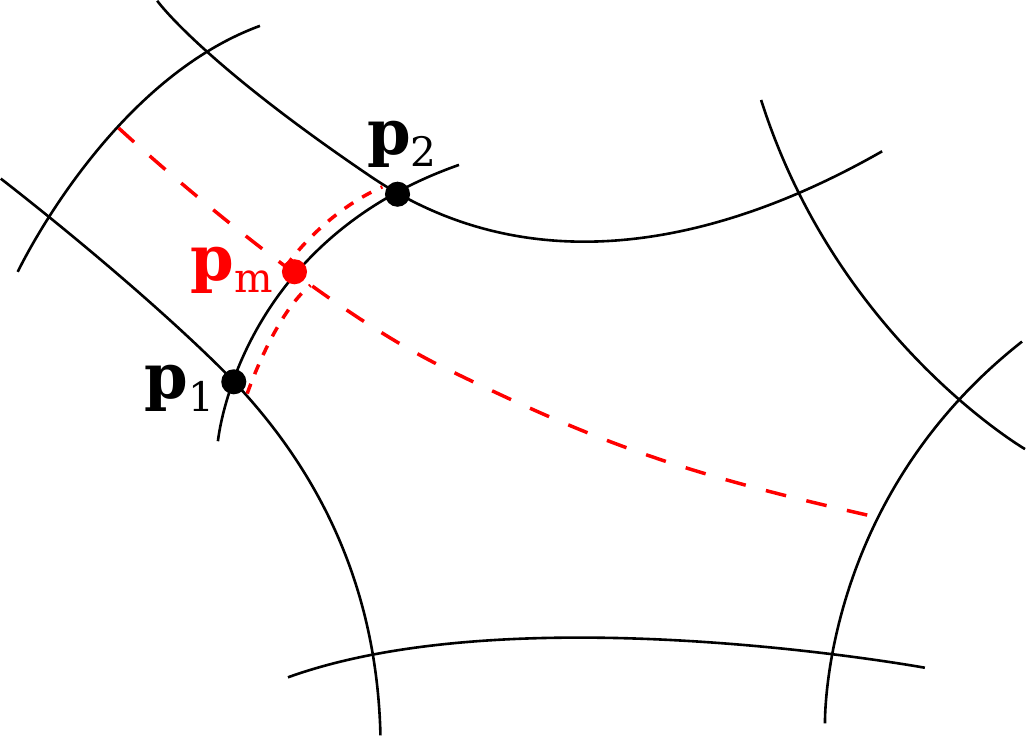}
      \caption{X-node}
      \label{fig:refinement-ex1}
    \end{subfigure}
    \hfill
    \begin{subfigure}{0.4\textwidth}
      \centering
      \includegraphics[width = \textwidth]{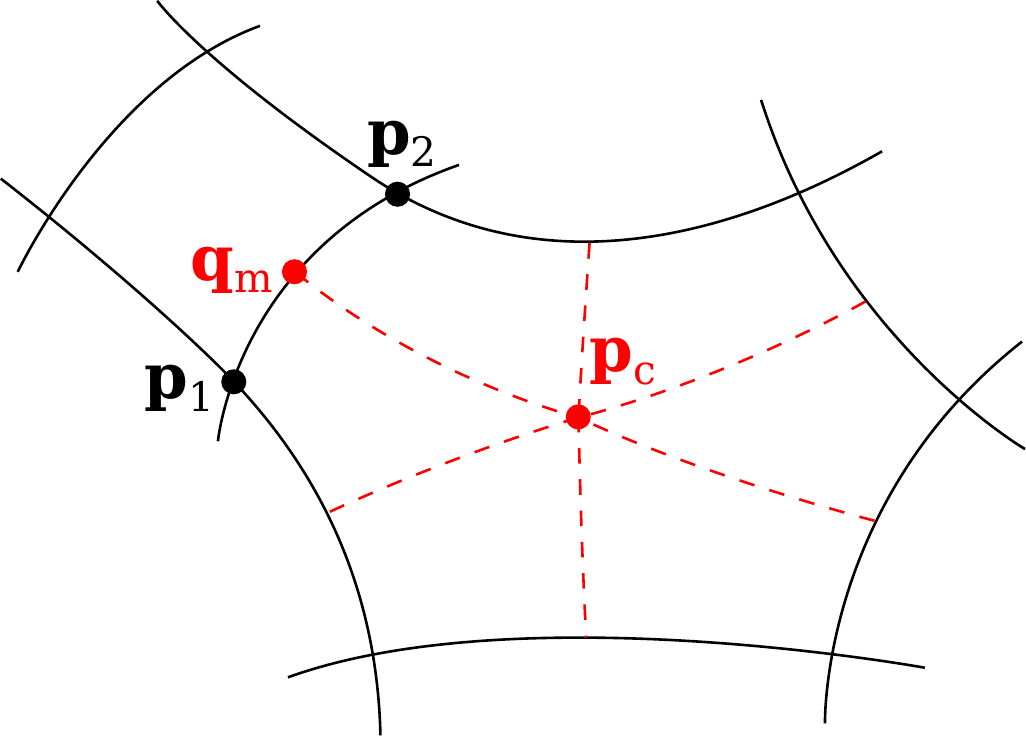}
      \caption{T-node}
      \label{fig:refinement-ex2}
    \end{subfigure}
    \hfill
  }
  \caption{Cases of adaptive refinement.}
  \label{fig:refinement}
\end{figure}

It is crucial, of course, to prevent the propagation of local
refinements over the full patchwork, and subdivide only where it is
necessary.  Fortunately, I-patches are well-suited for producing
T-nodes, as will be explained below.

Figure~\ref{fig:refinement-ex1} shows a boundary connecting
$\mathbf{p}_1$ and $\mathbf{p}_2$ that needs to be subdivided; a new
mesh point $\mathbf{p}_m$ is inserted, and four new boundaries are
created. This is an X-node, where we compute a new position, and a new
normal vector is \emph{taken from the mesh}. Figure~\ref{fig:refinement-ex2}
shows another configuration, where the 6-sided patch needs to be
centrally split at $\mathbf{p}_c$\rev{, which is the formerly described center
point of the original patch}. Here we wish to preserve the left
ribbon connecting $\mathbf{p}_1$ and $\mathbf{p}_2$, and let the two
adjacent subpatches of the subdivided patch \emph{inherit its midpoint}
$\mathbf{q}_m$ and the corresponding normal vector. In
other words, the original ribbon and bounding surface is transferred,
and $G^1$ continuity is naturally maintained. A related example will
be shown in the next section.

\section{EXAMPLES}
\label{sec:examples}

\begin{figure}[!ht]
  \centering
  \includegraphics[width = .55\textwidth]{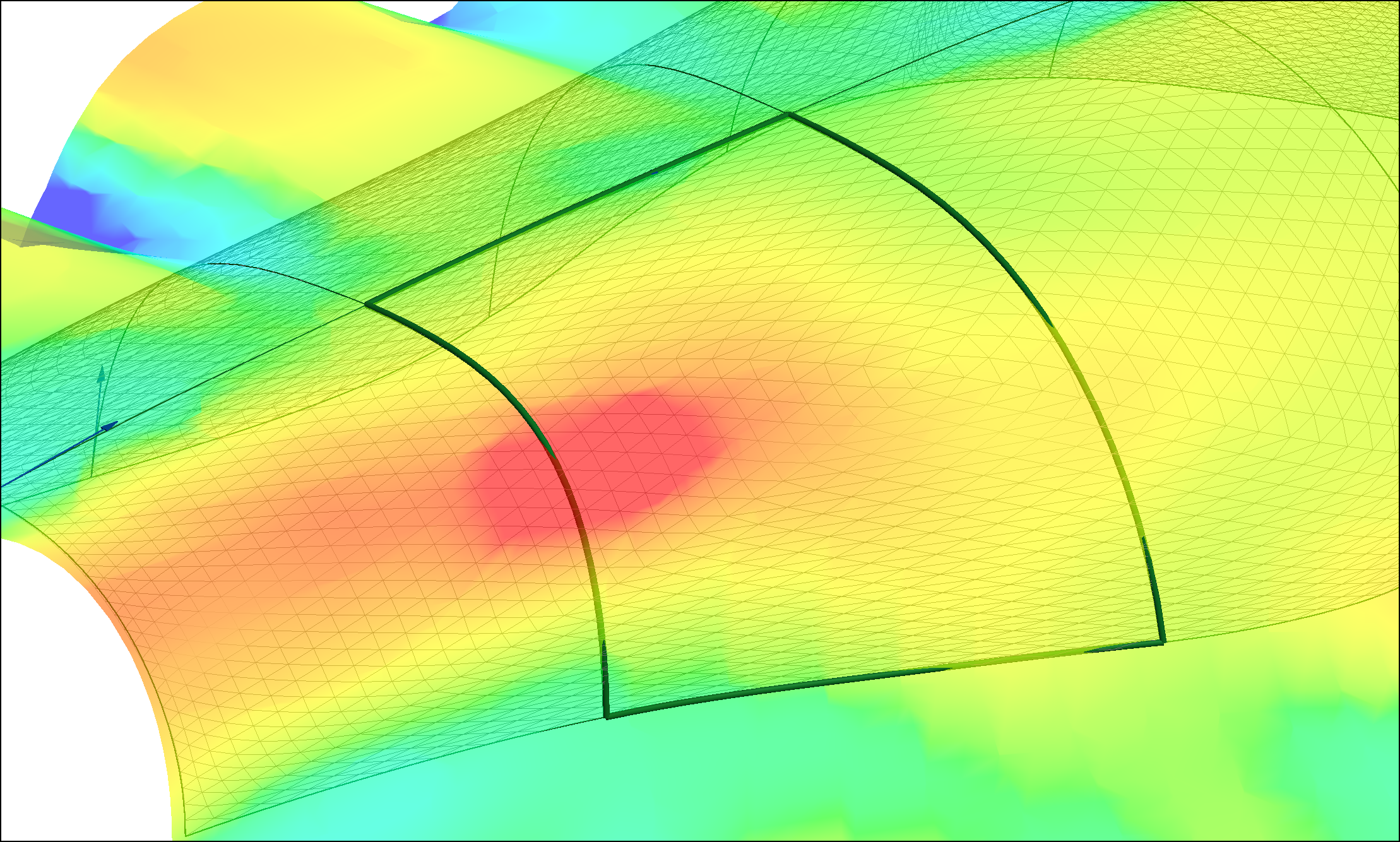}
  \caption{Natural extension of an I-patch.}
  \label{fig:faithful-b3}
\end{figure}

Our first example in Figure~\ref{fig:faithful-b3} illustrates that
I-patches naturally extend beyond the subspace determined by the
bounding surfaces. Here we rendered the curvature map of the patch
from Figure~\ref{fig:faithful-b1} using marching cubes.

\begin{figure}[!ht]
  {
    \hfill
    \begin{minipage}{.8\textwidth}
      \includegraphics[width = \textwidth]{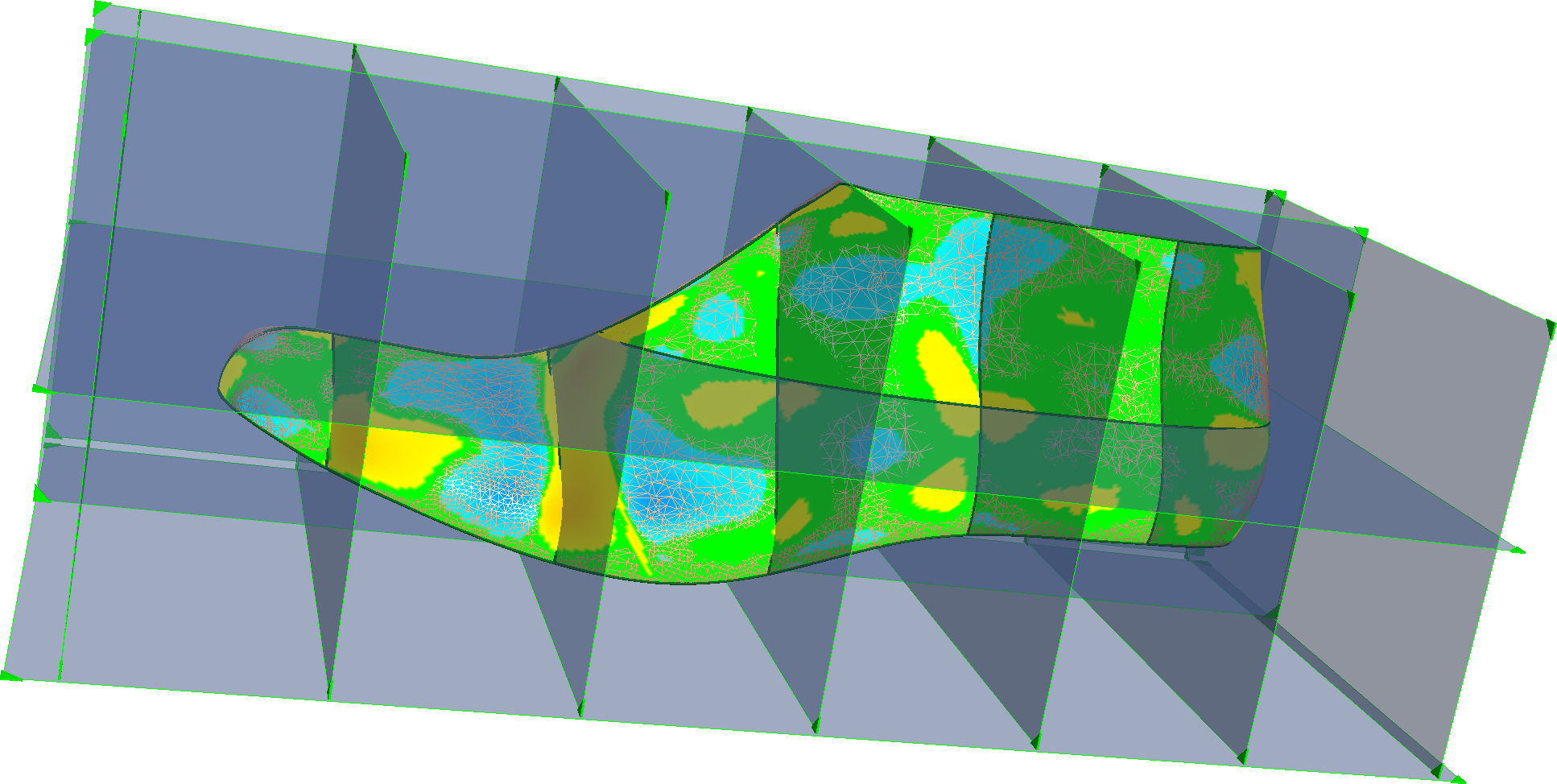}
    \end{minipage}
    \hfill
    \begin{minipage}{.1\textwidth}
      \includegraphics[width = \textwidth]{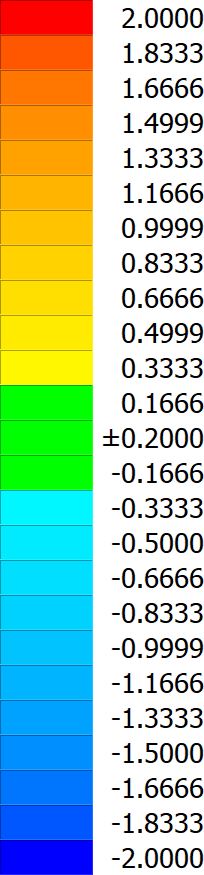}
    \end{minipage}
  }
  \caption{Cell-based subdivision on a model of a shoe last.
           \rev{The color legend shows absolute deviation values; the bounding box diagonal is 410 units long.}}
  \label{fig:cell}
\end{figure}

The second example in Figure~\ref{fig:cell} shows the deviation map of
a \emph{shoe last} model.  The network was created by a uniform cell
structure yielding two 3-sided, six 4-sided and two 5-sided
I-patches. The accuracy of the approximation---here and at the
forthcoming pictures---will be measured in percentages relative to the
diagonal of the bounding box.  The average \rev{(Euclidean)} deviation
from the mesh is 0.069\%, while the maximum is 0.35\%.

\begin{figure}[!ht]
  \begin{minipage}{.88\textwidth}
  {
    \hfill
    \begin{subfigure}{0.45\textwidth}
      \centering
      \includegraphics[width = \textwidth]{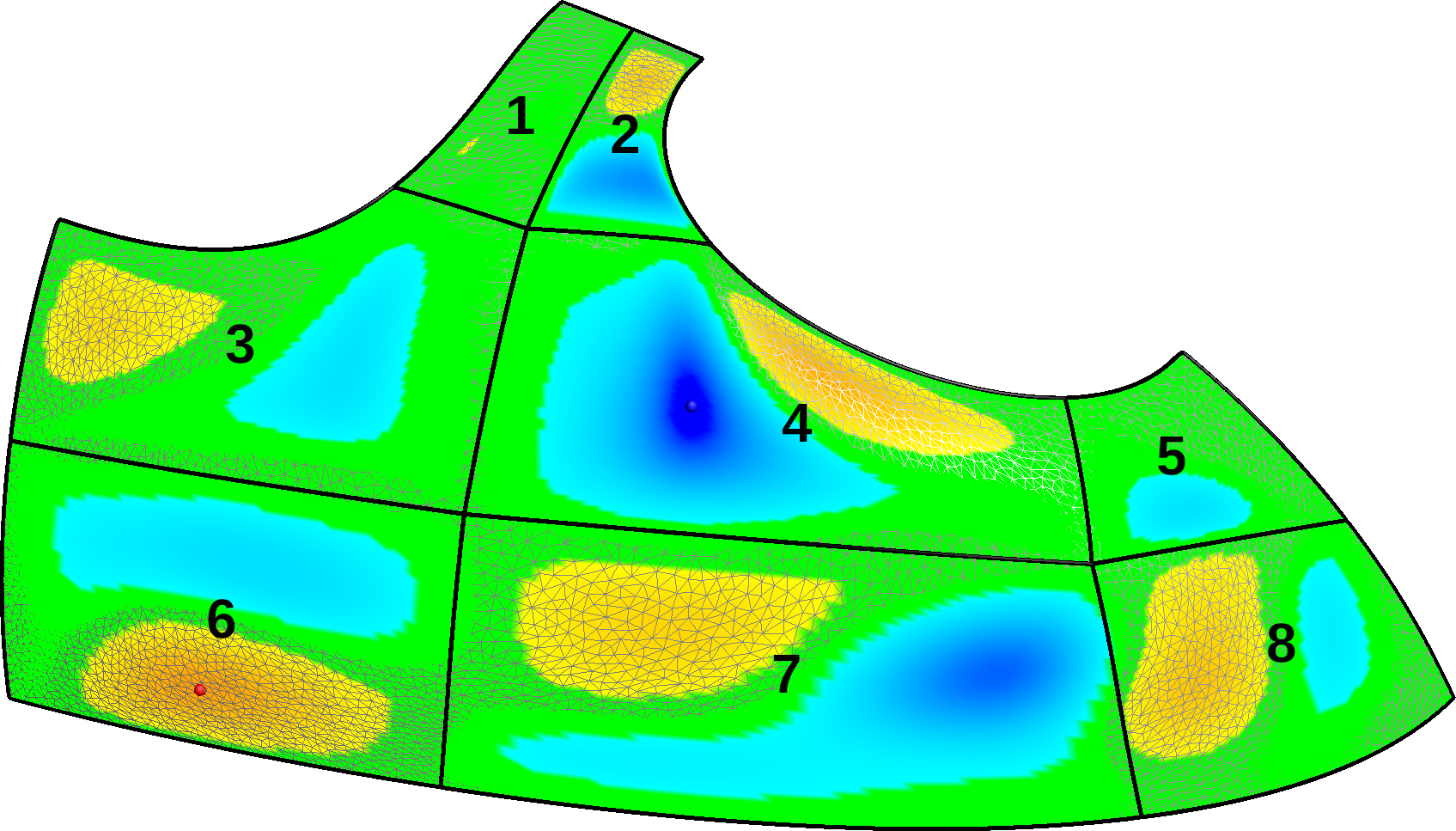}
      \caption{Original surface w/o optimization}
      \label{fig:refinement-base}
    \end{subfigure}
    \hfill
    \begin{subfigure}{0.45\textwidth}
      \centering
      \includegraphics[width = \textwidth]{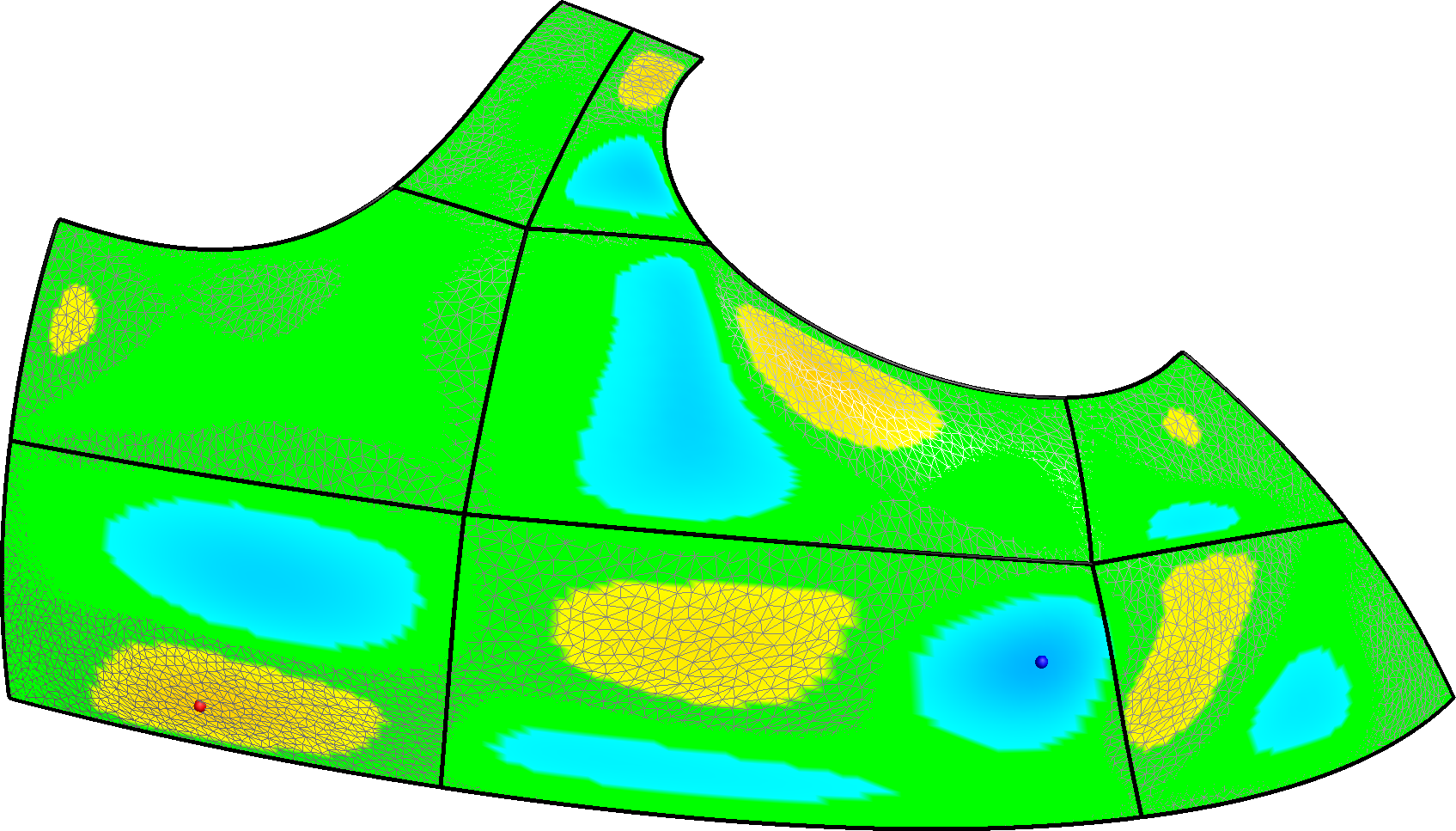}
      \caption{Original surface with optimization}
      \label{fig:refinement-base-opt}
    \end{subfigure}
    \hfill
  }

  {
    \hfill
    \begin{subfigure}{0.45\textwidth}
      \centering
      \includegraphics[width = \textwidth]{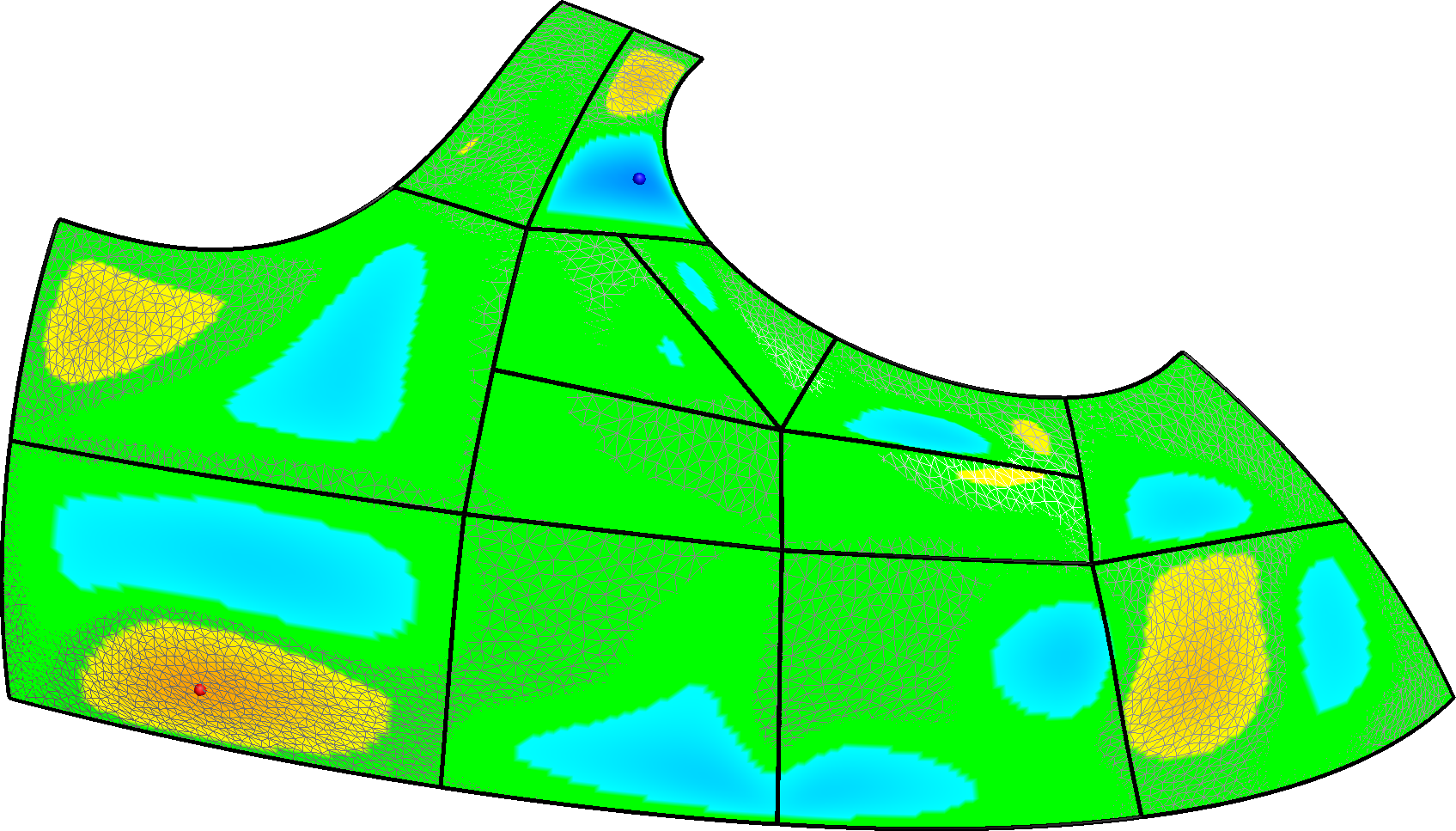}
      \caption{Subdivided surface w/o optimization}
      \label{fig:refinement-split}
    \end{subfigure}
    \hfill
    \begin{subfigure}{0.45\textwidth}
      \centering
      \includegraphics[width = \textwidth]{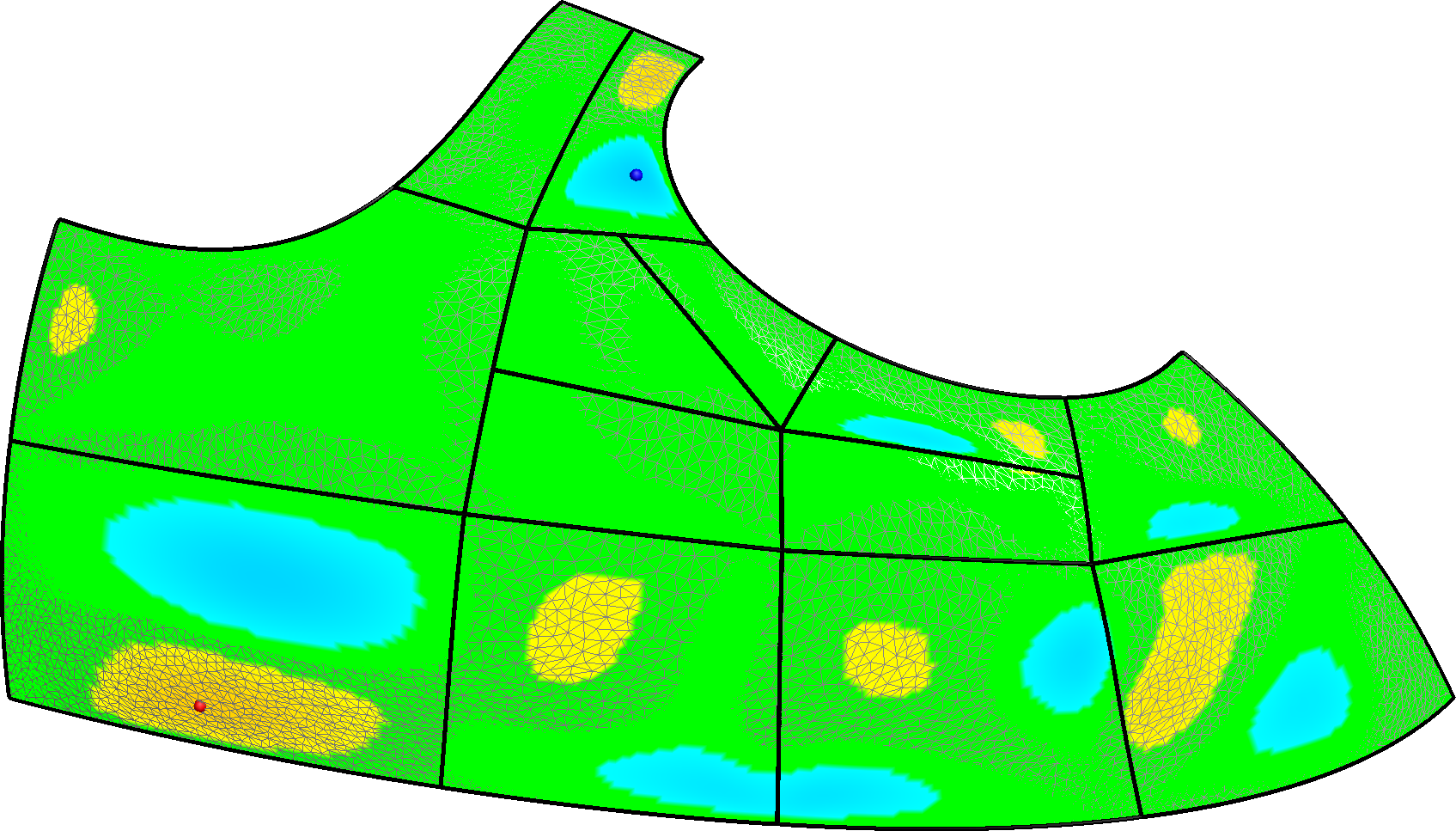}
      \caption{Subdivided surface with optimization}
      \label{fig:refinement-split-opt}
    \end{subfigure}
    \hfill
  }
  \end{minipage}
  \hfill
  \begin{minipage}{.1\textwidth}
    \includegraphics[width = \textwidth]{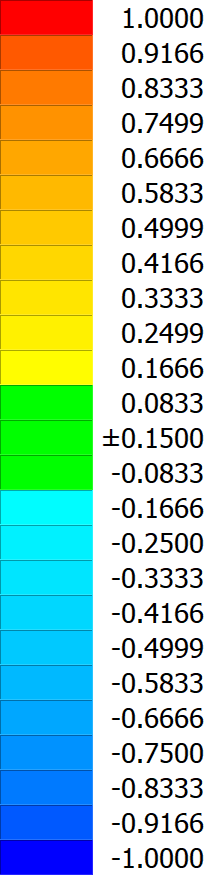}
  \end{minipage}
  \caption{Subdivision of a sheet metal part, with T-nodes and a split.
           \rev{The color legend shows absolute deviation values; the bounding box diagonal is 270 units long.}}
  \label{fig:vasalo}
\end{figure}

The third example (Fig.~\ref{fig:vasalo}) is a sheet metal part
composed of six 4-sided and two 5-sided patches, illustrating the
effect of optimization and refinement.
Figures~\ref{fig:refinement-base} and~\ref{fig:refinement-base-opt}
show two deviation maps, without and with optimization. 
Accuracies improve, as shown in Table~\ref{tab:optimization}.

Figures~\ref{fig:refinement-split} and~\ref{fig:refinement-split-opt}
show the deviation map of a refined structure---patch \#4 and patch
\#7 have been subdivided. \rev{As both of these patches are inaccurate,
their boundary is subdivided,} and a new X-node is inserted.
Boundaries 2--4, 3--4 and 4--5 are sufficiently accurate, so we retain
and inherit the ribbons from patches \#2, \#3 and \#5, and create
T-nodes accordingly, when patch \#4 is subdivided by a central
split. \rev{For simplicity's sake, planar bounding surfaces were
generated throughout this refinement.} The refined structure is more
accurate, and it can be further improved by optimization (see
Table~\ref{tab:optimization}).

\begin{figure}[!ht]
  \begin{minipage}{.88\textwidth}
  {
    \hfill
    \begin{subfigure}{0.45\textwidth}
      \centering
      \includegraphics[width = \textwidth]{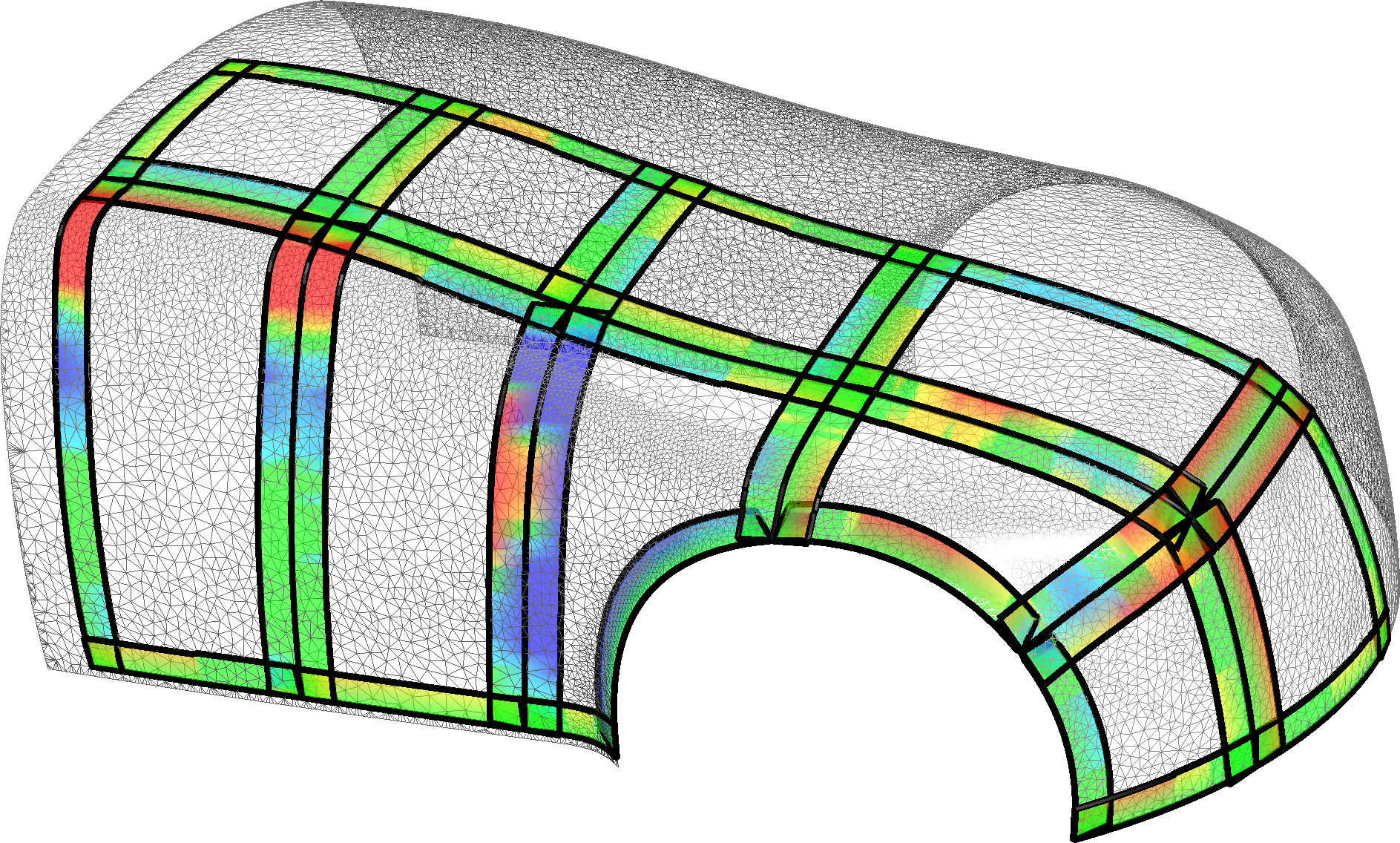}
      \caption{Initial ribbons}
      \label{fig:next30-initial-rib}
    \end{subfigure}
    \hfill
    \begin{subfigure}{0.45\textwidth}
      \centering
      \includegraphics[width = \textwidth]{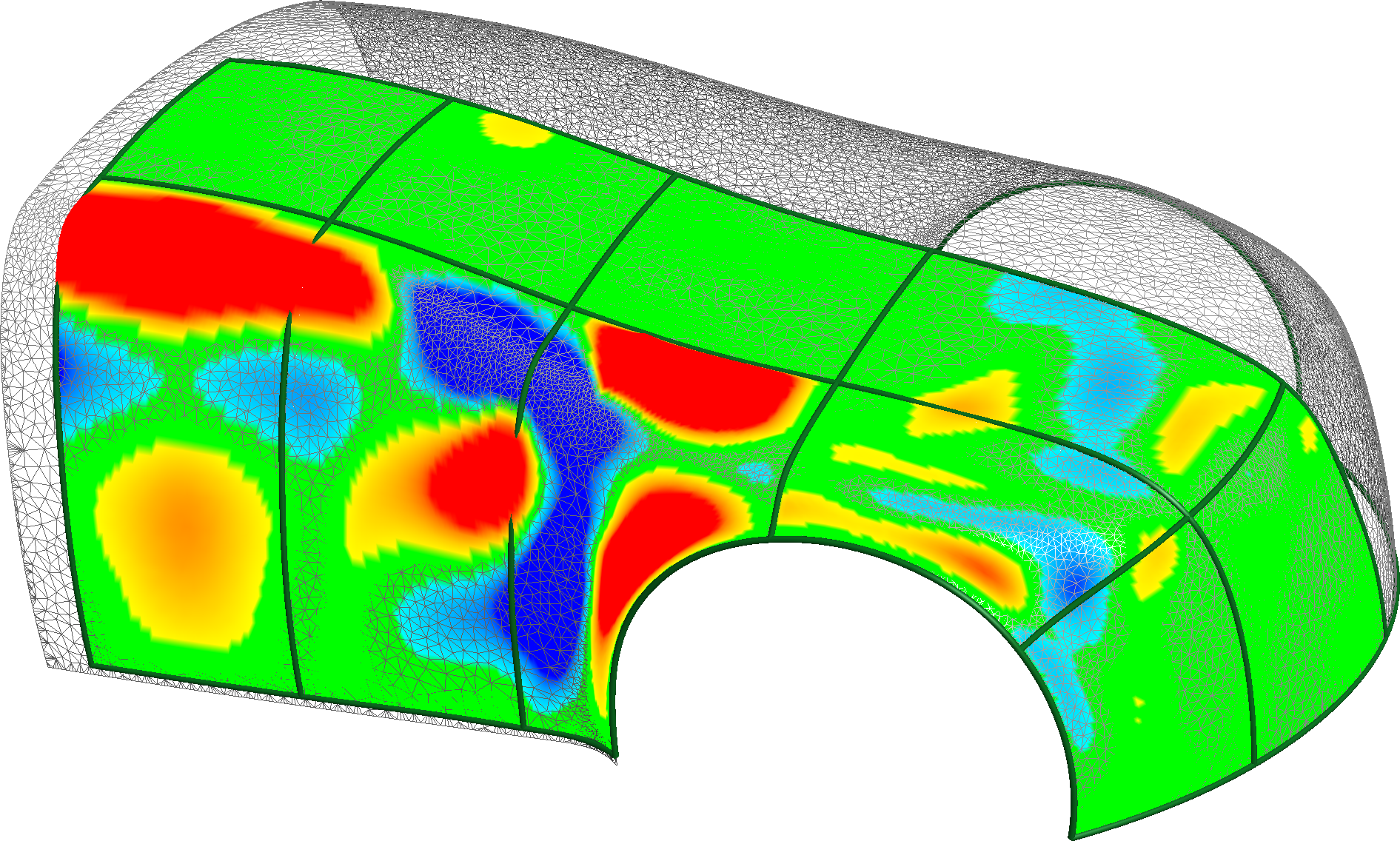}
      \caption{Initial patchwork}
      \label{fig:next30-initial}
    \end{subfigure}
    \hfill
  }

  {
    \hfill
    \begin{subfigure}{0.45\textwidth}
      \centering
      \includegraphics[width = \textwidth]{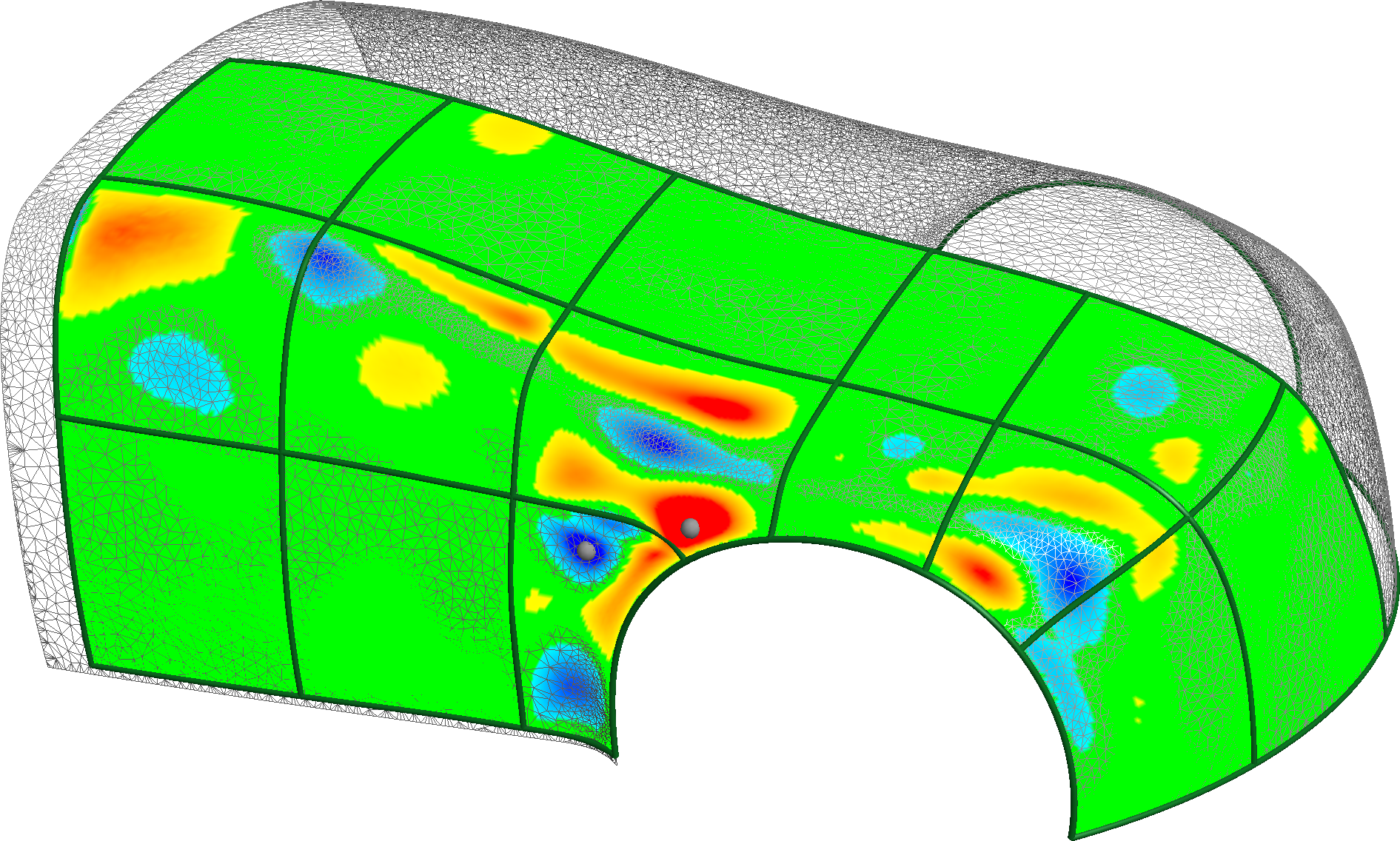}
      \caption{First refinement step}
      \label{fig:next30-step1}
    \end{subfigure}
    \hfill
    \begin{subfigure}{0.45\textwidth}
      \centering
      \includegraphics[width = \textwidth]{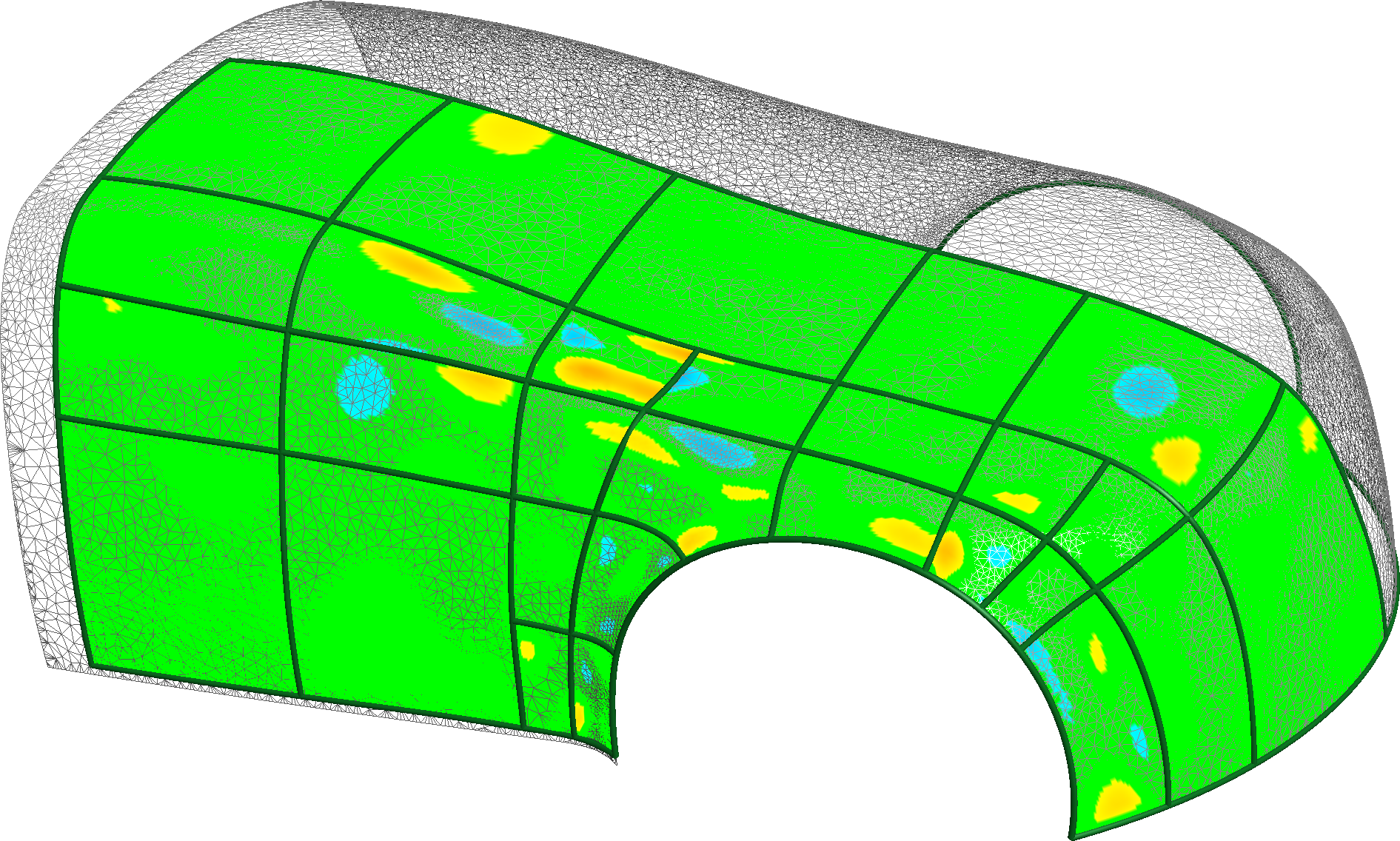}
      \caption{Second refinement step}
      \label{fig:next30-step2}
    \end{subfigure}
    \hfill
  }
  \end{minipage}
  \hfill
  \begin{minipage}{.1\textwidth}
    \includegraphics[width = \textwidth]{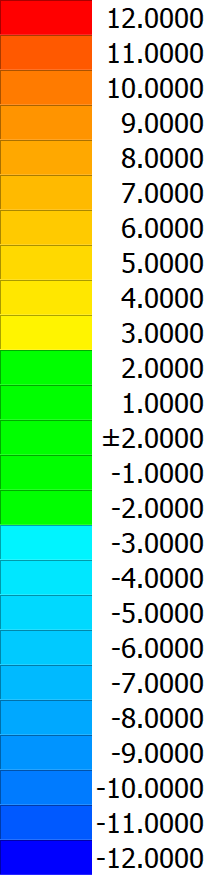}
  \end{minipage}
  \caption{Steps of adaptive refinement on a concept car model.
           \rev{The color legend shows absolute deviation values; the bounding box diagonal is 4400 units long.}}
  \label{fig:next30}
\end{figure}

Finally, Figure~\ref{fig:next30} shows the front part of a concept
car. The first image depicts the initial ribbons of a sparse network
that has been refined in two steps. The deviations decrease as the
number of I-patches increases, see Table~\ref{tab:next30}.

\begin{table}[!ht]
  \centering
  \begin{tabular}{c|c|c}
    Model & Average deviation & Maximum deviation\\
    \hline
    Original w/o optimization (Fig.~\ref{fig:refinement-base}) & 0.055\% & 0.399\% \\
    Original with optimization (Fig.~\ref{fig:refinement-base-opt}) & 0.035\% & 0.226\% \\
    Split w/o optimization (Fig.~\ref{fig:refinement-split}) & 0.041\% & 0.283\% \\
    Split with optimization (Fig.~\ref{fig:refinement-split-opt}) & 0.029\% & 0.175\%
  \end{tabular}
  \caption{Quantitative results corresponding to Figure~\ref{fig:vasalo}.
  }
  \label{tab:optimization}
\end{table}

\section{CONCLUSIONS}
\label{sec:conclusion}

We have shown that it is possible to approximate a free-form object by
a patchwork of smoothly connected, multi-sided implicit patches,
defined by corner points placed on a mesh, and a graph that determines
the connectivity between them. We have used a special class of
I-patches, constructed from implicit ribbon and bounding surfaces
determined solely by planes extracted from the mesh. The patches are
optimized by various fullness weights in a computationally efficient
manner, due to a faithful distance field that well approximates
Euclidean distances. Clearly, I-patches represent a more rigid variety
of shapes than their parametric counterparts, but nevertheless they
have several interesting geometric and computational properties.

Concerning future topics---we have used simple rules for defining the
ribbon and bounding surfaces, leaving many other options to be
explored in order to obtain better shapes and more complex patch
boundaries. We also consider determining the best possible
approximations for parametric patches. Efficient visualization of
I-patches using GPUs is also an interesting problem.

\section*{ACKNOWLEDGEMENTS}

\begin{table}[!ht]
  \centering
  \begin{tabular}{c|c|c|c}
    Model & \# of patches & Average deviation & Maximum deviation\\
    \hline
    Original~(Fig.~\ref{fig:next30-initial}) & 10 & 0.140\% & 3.000\% \\
    After 1 step~(Fig.~\ref{fig:next30-step1}) & 15 & 0.027\% & 0.300\% \\
    After 2 steps~(Fig.~\ref{fig:next30-step2}) & 28 & 0.016\% & 0.120\%
  \end{tabular}
  \caption{Quantitative results corresponding to Figure~\ref{fig:next30}.
  }
  \label{tab:next30}
\end{table}

This project has been supported by the Hungarian Scientific Research
Fund (OTKA, No.~124727), and the EFOP-3.6.1-16-2016-00014 project,
financed by the Ministry of Human Capacities of Hungary.  The
outstanding programming contribution of Gy\"orgy Karik\'o is highly
appreciated.

\bigskip
\orcid{\'Agoston Sipos,}{0000-0002-5562-2849}
\orcid{Tam\'as V\'arady,}{0000-0001-9547-6498}
\orcid{P\'eter Salvi,}{0000-0003-2456-2051}

\referenceSection
\bibliographystyle{CADA}
\bibliography{ipatch.bib}

\bigskip
\end{document}